\documentclass[twocolumn]{aastex62}

\usepackage{comment}
\usepackage{color}

\newcommand{\lsi}{\lambda_{\mathrm{si}}}
\newcommand{\lse}{\lambda_{\mathrm{se}}}
\newcommand{\opi}{\omega_{\mathrm{pi}}}
\newcommand{\ope}{\omega_{\mathrm{pe}}}
\newcommand{\Oi}{\Omega_{\mathrm{i}}}
\newcommand{\Oe}{\Omega_{\mathrm{e}}}

\newcommand{\thbn}{\theta_{\rm Bn}}

\newcommand{\ms}{M_\mathrm{s}}

\newcommand{\ok}{\textcolor{black}}
\newcommand{\rev}{\textcolor{black}}

\graphicspath{{./}{figures/}}

\received{\today}
\revised{\today}
\accepted{\today}
\submitjournal{ApJ}

\shorttitle{Low-Mach-Number Rippled Shocks}
\shortauthors{Kobzar et al.}

\begin{document}

\title{ELECTRON ACCELERATION AT RIPPLED LOW-MACH-NUMBER SHOCKS \\
       IN HIGH-BETA COLLISIONLESS COSMIC PLASMAS}

\correspondingauthor{Oleh Kobzar}
\email{oleh.kobzar@oa.uj.edu.pl}

\author[0000-0001-6956-5884]{Oleh Kobzar}
\affiliation{Astronomical Observatory of the Jagiellonian University, PL-30244 Krak{\' o}w, Poland}
\affiliation{Faculty of Materials Engineering and Physics, Cracow University of Technology, PL-30084 Krak{\' o}w, Poland}

\author{Jacek Niemiec}
\affiliation{Institute of Nuclear Physics Polish Academy of Sciences, PL-31342 Krak{\' o}w, Poland}

\author{Takanobu Amano}
\affiliation{Department of Earth and Planetary Science, the University of Tokyo, 7-3-1 Hongo, Bunkyo-ku, Tokyo 113-0033, Japan}

\author{Masahiro Hoshino}
\affiliation{Department of Earth and Planetary Science, the University of Tokyo, 7-3-1 Hongo, Bunkyo-ku, Tokyo 113-0033, Japan}

\author{Shuichi Matsukiyo}
\affiliation{Faculty of Engineering Sciences, Kyushu University, 6-1 Kasuga-Koen, Kasuga, Fukuoka, 816-8580, Japan}

\author{Yosuke Matsumoto}
\affiliation{Department of Physics, Chiba University, 1-33 Yayoi-cho, Inage-ku, Chiba 263-8522, Japan}

\author{Martin Pohl}
\affiliation{Institute of Physics and Astronomy, University of Potsdam, 14476 Potsdam-Golm, Germany}
\affiliation{DESY, Platanenallee 6, 15738 Zeuthen, Germany}


\begin{abstract}

Using large-scale fully-kinetic two-dimensional particle-in-cell simulations, we investigate the effects of shock rippling on electron acceleration at low-Mach-number shocks propagating in high-$\beta$ plasmas, in application to merger shocks in galaxy clusters. We find that the electron acceleration rate increases considerably when the rippling modes appear. The main acceleration mechanism is stochastic shock-drift acceleration, in which electrons are confined at the shock by pitch-angle scattering off turbulence and gain energy from the motional electric field. The presence of multi-scale magnetic turbulence at the shock transition and the region immediately behind the main shock overshoot is essential for electron energization. Wide-energy non-thermal electron distributions are formed both upstream and downstream of the shock. The maximum energy of the electrons is sufficient for their injection into diffusive shock acceleration. We show for the first time that the downstream electron spectrum has a~power-law form with index 
\rev{$p\approx 2.5$}, in agreement with observations.


\end{abstract}


\section{Introduction}
\label{sec:intro}

Merger shocks at outskirts of galaxy clusters have been detected through X-ray observations \citep[e.g.,][]{Markevitch-02,Russell-10,2017A&A...600A.100A} and show radio synchrotron emission from relativistic electrons in the so-called radio relics \citep[e.g.,][]{Willson-70, Fujita-01,Govoni-04,van-Weeren-10,Lindner-14}.
These electrons are presumably accelerated at large-scale shock fronts, that are also thought to be possible sources of ultra-high-energy cosmic rays (UHECRs) with energy exceeding $10^{18}$ eV, though $\gamma$-ray emission from galaxy clusters, which would be a unique signature of CR protons, has not been detected so far \citep[see, e.g.,][]{Brunetti-14}. 
The connection of radio relics to shocks suggests electron production via diffusive shock acceleration (DSA), also known as the first-order Fermi process \citep[e.g.,][]{Drury-83, Blandford-87}. 
In this process particles gain their energies in repetitive interactions with the shock front. The critical unresolved problem in DSA theory is the particle injection. DSA works only for particles that have Larmor radii much larger than the internal shock width, typically a few gyroradii of thermal ions. 
Therefore, both electrons and ions need to be pre-accelerated to suprathermal momenta, $p_{\rm inj} \sim {\rm a\, few\,} p_{\rm th, p}$, where $p_{\rm th, p}$ is the momentum of postshock thermal ions. Achieving $p_{\rm inj}$ is more difficult for electrons than for protons, on account of their lower mass and smaller Larmor radii. Electron pre-acceleration thus likely arises from other interactions than those providing ion acceleration. This is known as the electron injection problem.

Merger shocks have very low sonic Mach numbers, $\ms\lesssim 4$, and propagate in the hot intracluster medium (ICM), in which the plasma beta (a ratio of thermal to magnetic pressure) is high, $\beta\gg 1$. Particle acceleration is poorly known in this regime. 
Electron acceleration at low-Mach-number high-$\beta$ collisionless shocks has recently been studied with kinetic particle-in-cell (PIC) simulations. 
One-dimensional (1D) simulations by \cite{Matsukiyo-11} and later two-dimensional (2D) studies by \cite{Park-12, Park-13} demonstrated that in such shocks electrons can be efficiently energized via shock drift acceleration (SDA). 
In this process, particles drift along the shock surface due to the magnetic field gradient at the shock, and gain their energies from the shock motional electric field \citep{Wu-84, Krauss-Varban-89, Ball-01, Mann-06, Park-13}:

\begin{equation}
\Delta \gamma_{\mathrm{SDA}} = \frac{-e}{m_{\mathrm{e}} c^2} \int E_{z}\,dz
\approx \frac{-e}{m_{\mathrm{e}} c^2}\,E_{z}^{\mathrm{up}}\,\Delta z~,
\label{eq:SDA}
\end{equation}
where $\gamma$ is the Lorentz factor, $E_{z}^{\mathrm{up}}$ is the upstream (motional) electric field, and $\Delta z$ is the path-length of the particle drift.
In conditions of high plasma $\beta$ and at oblique \textit{subluminal} shocks some of the SDA-accelerated electrons can be reflected at the shock 
and form non-equilibrium velocity distribution in the foreshock region that leads to instabilities which generate waves.
It was suggested by \cite{Matsukiyo-11} that electrons can be scattered off these waves back to the shock and undergo further energization. This scenario has been confirmed in 2D simulations by \citet{Guo-14a, Guo-14b}, which showed that upstream electron scattering allows for multiple SDA cycles resembling a sustained DSA process.

The waves providing electron scattering have been identified in \citet{Guo-14a, Guo-14b} as an oblique mode of the electron firehose instability (EFI). This instability can be driven by the electron temperature anisotropy that is effectively created when the reflected electrons stream along the mean magnetic field \citep[e.g.][see also \citet{Kim-20} for the electron beam driven EFI modes]{Li-00, Gary-03, Camporeale-08}. 
Systematic investigations indicated that this mechanism of wave generation and electron scattering works at low-Mach-number shocks for temperatures relevant for galaxy clusters and a wide range of magnetic-field inclination angles, $\thbn$, and in particular in high beta plasmas, $\beta \gtrsim 20$ \citep{Guo-14b}.
For shock obliquities enabling a large flux of reflected electrons and hence a strong temperature anisotropy, non-thermal electrons were found with a power-law energy distribution, $dn/dE_{\mathrm{kin}} \propto E_{\mathrm{kin}}^{-p}$, with a slope $p\simeq 2.4$ that corresponds to the spectral index of radio synchrotron emission $\alpha = -0.7$, compatible with observations \citep[e.g.,][]{van-Weeren-10}. However, such distributions were found in the \textit{upstream} spectra only, and the downstream spectra remained approximately thermal.
Most recently, \citet{Kang-19} showed that electron pre-acceleration via SDA can occur only at shocks exceeding the so-called EFI-critical Mach number, $M_{\mathrm{ef}}^{*} \approx 2.6$, 
which is higher than the critical Mach number $M_{\rm crit}\approx 1.26$ that one derives from the MHD jump conditions in low $\beta$ shocks. This suggests that shocks with $\ms\lesssim 2.3$ cannot accelerate electrons. Moreover, even at supercritical shocks with $\ms\gtrsim M^*_{\rm ef}$ electrons may not reach a sufficiently high energy to be injected to DSA, because EFI was observed to saturate and did not generate long-wavelength modes. 

The PIC studies reported above used relatively narrow simulation boxes that resolve only electron-scale structures. Ion-scale fluctuations, e.g., in the form of the shock corrugations, have not been accounted for. The first large-scale 2D simulation resolving the multi-scale shock structure has been reported by \citet{Matsukiyo-15} for the shock with $M_{\mathrm{s}} = 2.6$ and $\beta = 3$ that was studied earlier with 1D simulations \citep{Matsukiyo-11}, showing efficient SDA. Shock rippling was observed to spawn local regions with weaker magnetic field along the corrugated shock. Most electrons encounter a weak-field region during their SDA interaction with the shock, which drastically increases the likelihood of their transmission to the downstream region and reduces the probability of reflection.
Some non-thermal electrons can still be found at the shock, but they result from local wave-particle interactions in the shock transition.

The origin of shock rippling in the simulations by \citet{Matsukiyo-15} is considered to be the downstream ion temperature anisotropy provided by gyrating shock-reflected ions that are advected back downstream.
In this case the Alfv\'{e}n ion cyclotron (AIC) instability can be triggered. With increasing plasma beta, the temperature anisotropy becomes smaller, the growth rate of the AIC instability is lower, and the rippling modes have larger wavelengths. 
It was estimated that the wavelength of the ripples in the high-beta simulations ($\beta \geq 20$) by \citet{Guo-14a,Guo-14b} and \citet{Kang-19} is much larger than the transverse system size they used, so that the modes could not be captured. 

In the present work we investigate the effects of shock rippling on electron injection at low-Mach-number shocks in high-beta plasma. Our large-scale 2D PIC simulations are performed in a parameter regime in which Fermi-like acceleration can operate.
The simulation setup is described in Section~\ref{sec:setup}. The evolution of the shock structure and electron energy distribution are considered in Sections \ref{sec:evolution} and \ref{sec:distr}, correspondingly. In Section~\ref{sec:accel} we discuss the micro-physics of the electron acceleration processes, and we summarize the results in Section~\ref{sec:sum}. 
Preliminary results of these studies have been presented in \citet{2019ICRC...36..368N}.

\section{Simulation setup and parameters}
\label{sec:setup}

\begin{figure}
\centering
\includegraphics[width=0.99 \linewidth, clip]{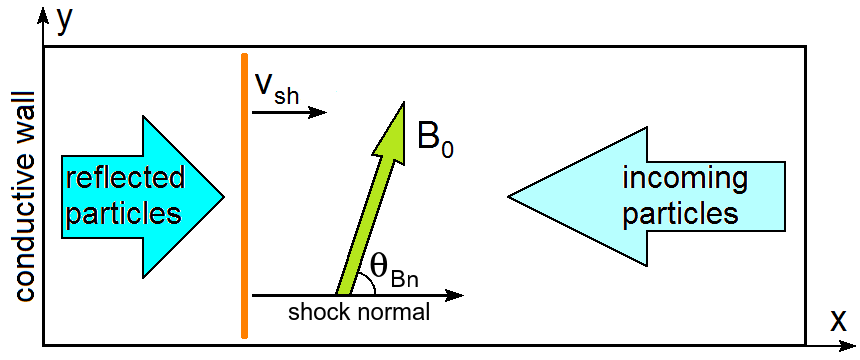}
\caption{The simulation setup with conductive reflecting wall. The motional electric field $\bm{E}_0 = -[\bm{v}_{0} \times \bm{B}_{0}]$ is directed out of 2D simulation plane in $+z$-direction.}
\label{fig:box}
\end{figure}

We use a modified version of the relativistic electromagnetic PIC code TRISTAN \citep{Buneman-93} with MPI-based parallelization \citep{Niemiec-08} and the option to follow selected individual particles. We apply a 2D3V simulation model that utilizes a two-dimensional spatial grid in the $x - y$ plane and follows all three components of particle momenta and electromagnetic fields. The simulation setup is shown in Figure~\ref{fig:box}.
An electron-ion plasma beam is injected at the right side of the simulation box to flow in the negative $x$-direction with bulk speed $v_{0}$. After reflection off the conductive wall at the left boundary, the beam interacts with the incoming plasma and forms a shock that propagates in the $+x$-direction with the speed $v_{\mathrm{sh}}$. The right ($x$-) boundary is open, and we apply periodic boundary conditions in $y$ direction.

The injected plasma carries a large-scale magnetic field, $\bm{B}_{0}$, which lies in the simulation plane at an angle $\theta_{\mathrm{Bn}} = 75^{\circ}$ to the shock normal.
We therefore study a quasi-perpendicular \textit{subluminal} shock, as the critical superluminality angle is $\theta_{\mathrm{Bn,cr}} = \cos^{-1}{(v_{\mathrm{sh}}^{\mathrm{up}}/c)} \approx 81.4^{\circ}$.
Together with the magnetic field, a motional electric field $\bm{E}_0 = -[\bm{v}_{0} \times \bm{B}_{0}]$ is initialized that is directed out-of-plane, $\bm{E}_{0}=E_{0z}\bm{\hat{z}}$.

The simulation parameters have been chosen to represent physical conditions typical for shock waves in ICM. The bulk plasma flow velocity is $v_{0} = 0.1\,c$, where $c$ is the speed of light. The electrons and ions are initially in thermal equilibrium with 
temperatures $T_{\mathrm{e}} = T_{\mathrm{i}}  \rev{= T_0} \approx 5 \cdot 10^{8}\, \mathrm{K} = 43\, \mathrm{keV}/k_{\mathrm{B}}$. 
With these parameters the sonic Mach number of the shock measured in the upstream plasma rest frame is $M_{\mathrm{s}} \equiv v_{\mathrm{sh}}^{\mathrm{up}}/c_{\mathrm{s}} = 3$, where the sound speed $c_{\mathrm{s}} = \sqrt{2 \Gamma k_{\mathrm{B}} T_{\mathrm{i}} / m_{\mathrm{i}}}$, and $\Gamma$ is the adiabatic index. 
The Alfv\'{e}nic Mach number is $M_{\mathrm{A}} \equiv v_{\mathrm{sh}}^{\mathrm{up}}/v_{\mathrm{A}} \approx 6.1$, where $v_{\mathrm{A}} = B_{\mathrm{0}} / \sqrt{\mu_{\mathrm{0}} (N_{\mathrm{i}} m_{\mathrm{i}} + N_{\mathrm{e}} m_{\mathrm{e}}) }$ is the Alfv\'{e}n velocity, $\mu_{\rm 0}$ is the vacuum permeability, and $N_{\mathrm{i}}$ and $N_{\mathrm{e}}$ are the upstream ion and electron number densities.
The total plasma beta, 
$$ \beta \equiv \frac{p_{\mathrm{th}}}{p_{\mathrm{m}}}= \frac{2\mu_0(N_{\mathrm{e}}+N_{\mathrm{i}})k_{\mathrm{B}}\rev{T_0}}{B_0^2}=5 ~,$$
is equally carried by electrons and ions, $\beta_{\mathrm{e}} = \beta_{\mathrm{i}} = 2.5$.
This value is lower than that in earlier simulations in which the EFI is efficiently excited \citep{Guo-14a, Guo-14b}, but this choice is  necessary to fit the ion-scale rippling modes into the simulation box and to follow the long-term evolution of the system. For the same reason we apply a reduced ion-to-electron mass ratio, $m_{\mathrm{i}}/m_{\mathrm{e}} = 100$.
We expect the wavelength of the rippling modes to be in the range $15\, \lsi\lesssim \lambda_{\mathrm{rippl}} \lesssim 20\, \lsi$.

\begin{figure*}[t!]
\centering
\includegraphics[width=0.49 \linewidth, clip]{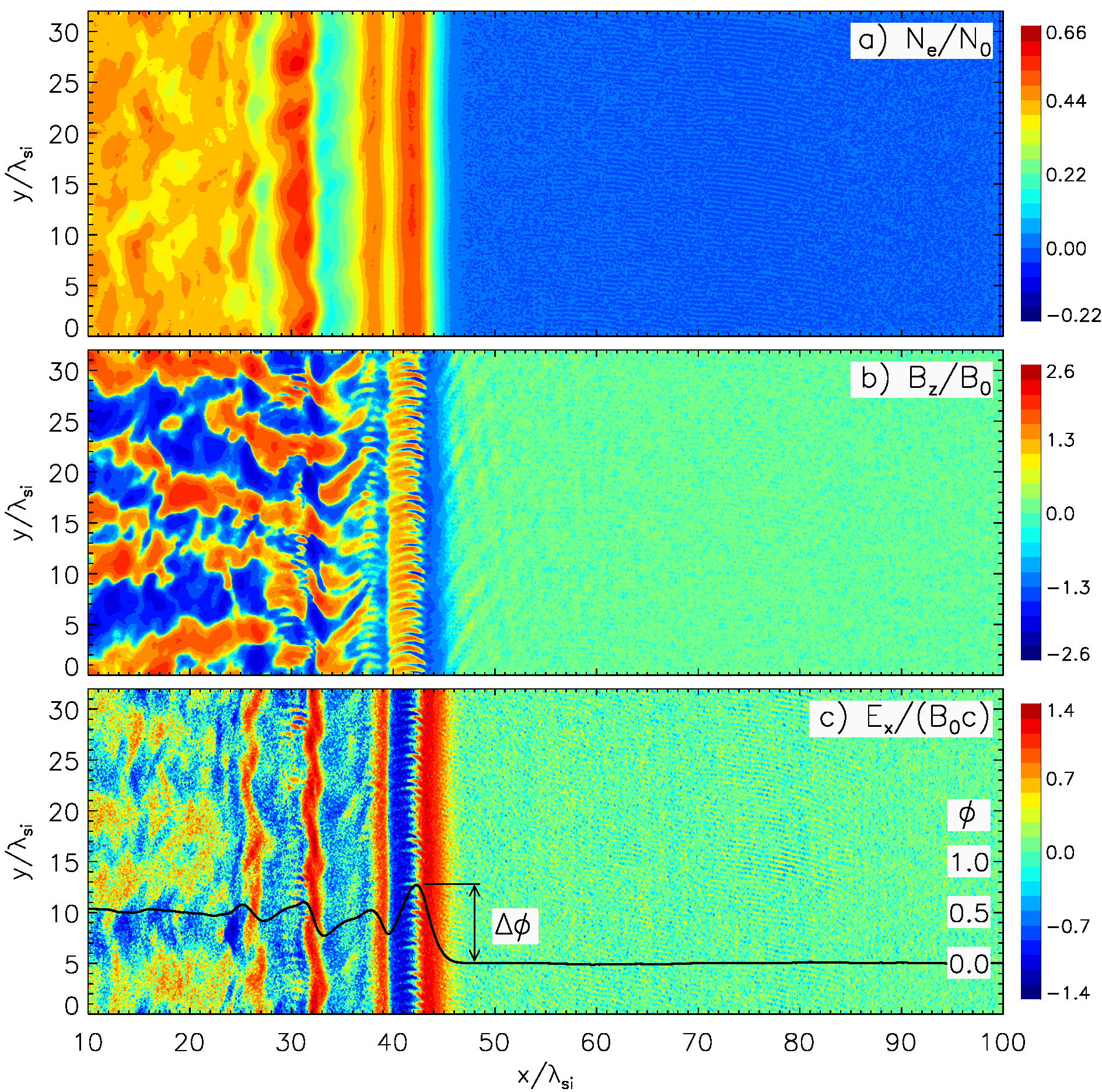}
\includegraphics[width=0.49 \linewidth, clip]{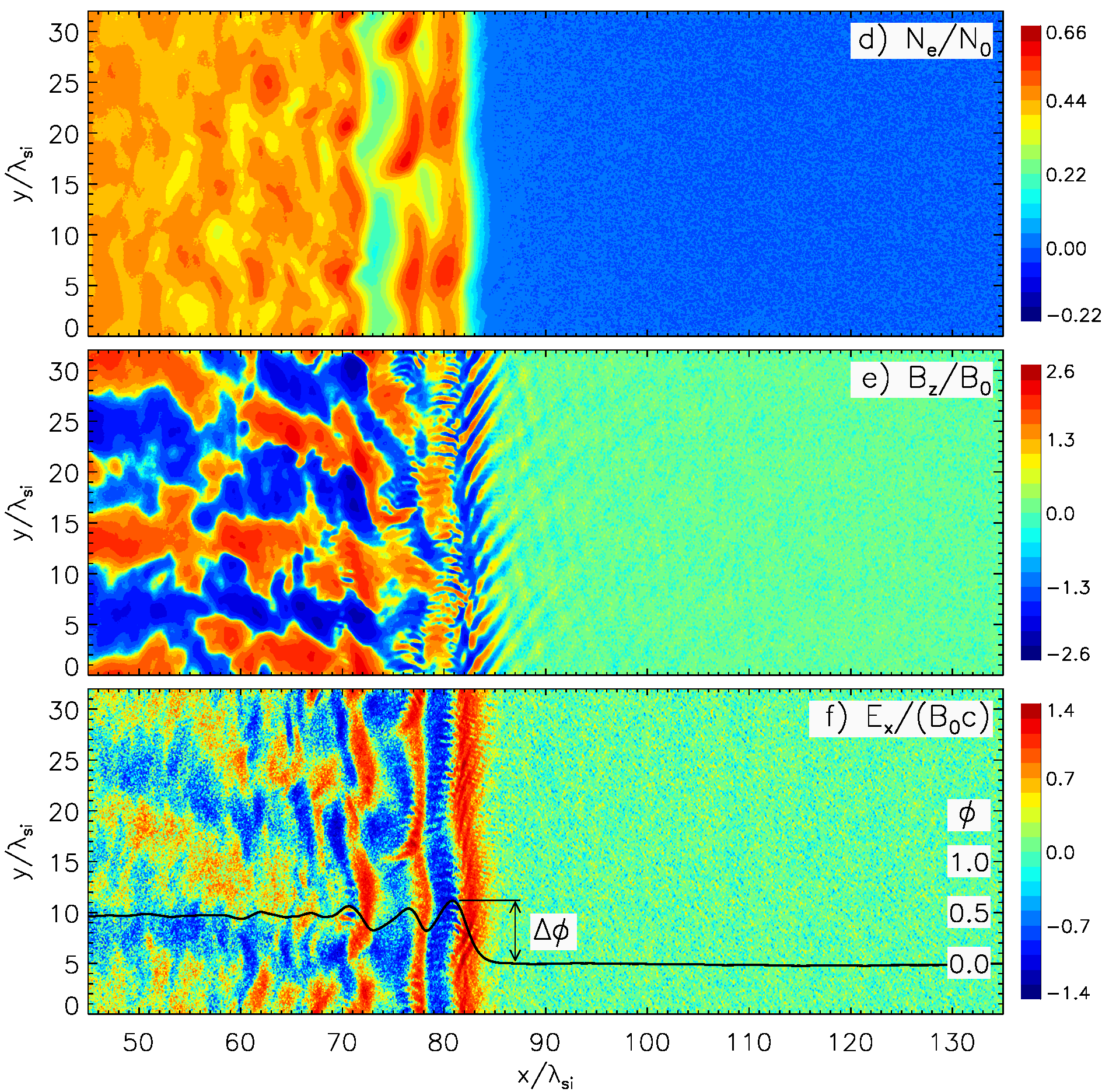}
\caption{Structure of the \textit{laminar} (left, $\Oi t = 18$) and \textit{rippled} (right, $\Oi t = 36$) shock. Shown are distributions of the normalized electron number density, $N_{\mathrm{e}}/N_{0}$ (Panels \textit{a}, \textit{d}), the normalized magnetic field, $B_{z}/B_{0}$ (\textit{b}, \textit{f}), and the normalized electric field, $E_{x}/(B_{0}c)$ (\textit{c}, \textit{f}). The density maps have logarithmic scaling. The scaling for magnetic and electric fields is also logarithmic, but sign-preserving, and, e.g., for $B_{z}$ it is: $\mathrm{sgn}(B_{z}) \cdot \{2+\log[\max(|B_{z}|/B_{0},10^{-2})]\}$. The level of "0" on the color scale hence corresponds to $|B|/B_{0} \le 10^{-2}$, and likewise for the electric field. In panels (\textit{c}) and (\textit{f}), solid curves show the normalized, $y$-averaged 
electric potential energy, $\phi$ (Eq.~\ref{eqphi}),
calculated in the shock rest frame. The cross-shock potential amplitude in the shock rest frame, $\Delta\phi$, is marked with arrows. }
\label{fig:maps}
\end{figure*}

Convergence tests suggest that it is sufficient to inject
20 particles per cell per species in the upstream plasma and to set the electron skin depth to
$\lse \equiv c/\ope = 15 \mathrm{~cells}$, where 
$\ope = \sqrt{e^{2} N_{\mathrm{e}} / \varepsilon_{\mathrm{0}} m_{\mathrm{e}}}$ is the electron plasma frequency, with the electron charge, $e$, and the vacuum permittivity, $\varepsilon_0$. The ion skin depth, $\lsi = \lse \sqrt{m_{\mathrm{i}}/m_{\mathrm{e}}} = 150 \mathrm{~cells}$, is the main unit of length in our simulations. Time is given in units of the upstream ion cyclotron frequency, $\Oi = eB_{\mathrm{0}}/m_{\mathrm{i}}$.
The maximum simulation time is $t_{\mathrm{max}}\Oi \approx 78$.
The time step is $\delta t = 1/(30\,\ope)=1/(1.225 \cdot 10^{4}\,\Oi)$.
The transverse size of the simulation box is $L_{y} = 320\, \lse = 32\, \lsi$. 
Fresh particles are added at a \emph{moving injection layer} that recedes from the shock, so that the simulated plasma contains all reflected particles. The final box length is 
$L_{x} \approx 4000\, \lse = 400\, \lsi$.

\section{Evolution of the shock}
\label{sec:evolution}

\subsection{Shock Structure}
\label{subsec:structure}

\begin{figure*}
\centering
\includegraphics[width=0.49 \linewidth, clip]{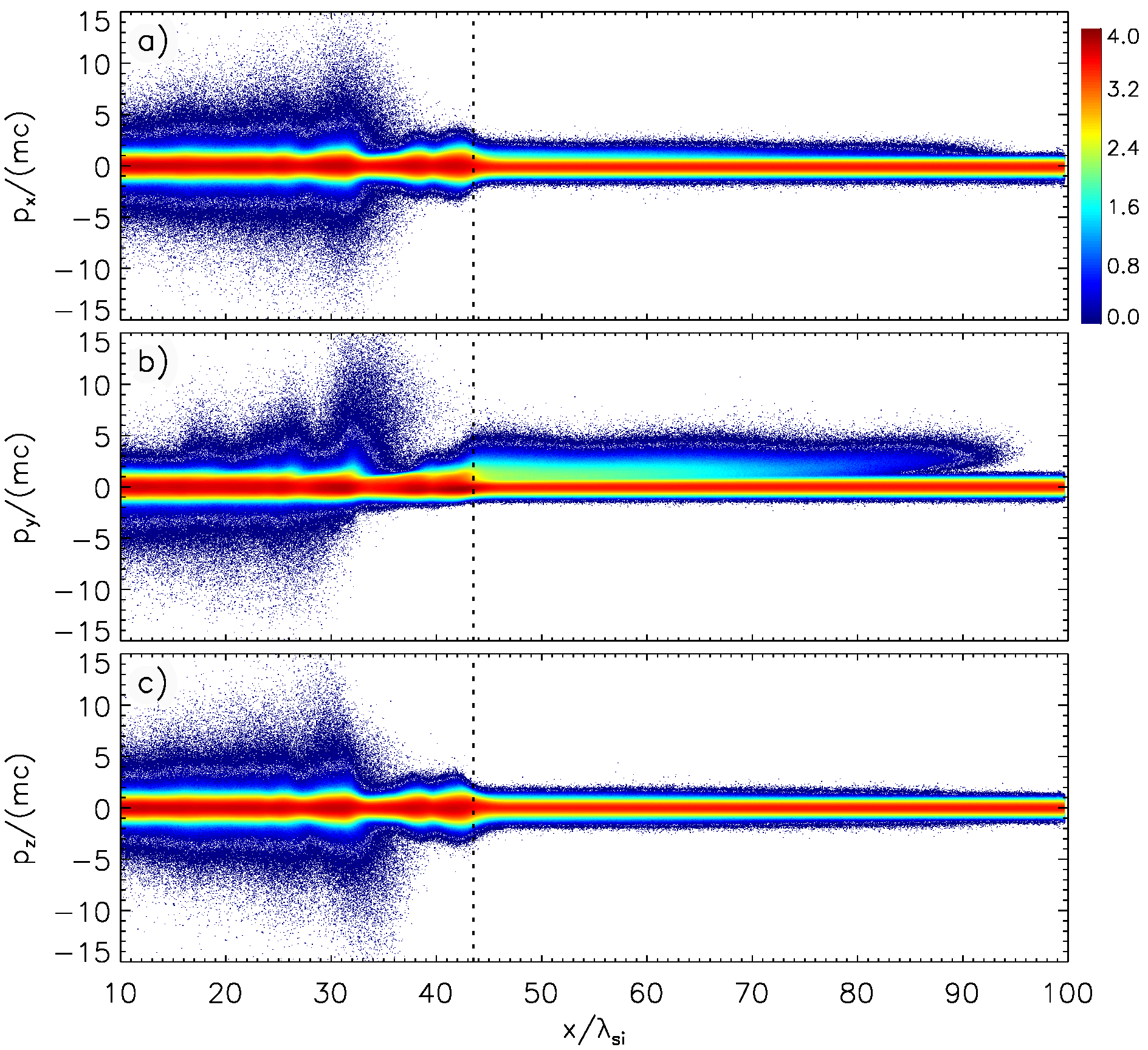}
\includegraphics[width=0.49 \linewidth, clip]{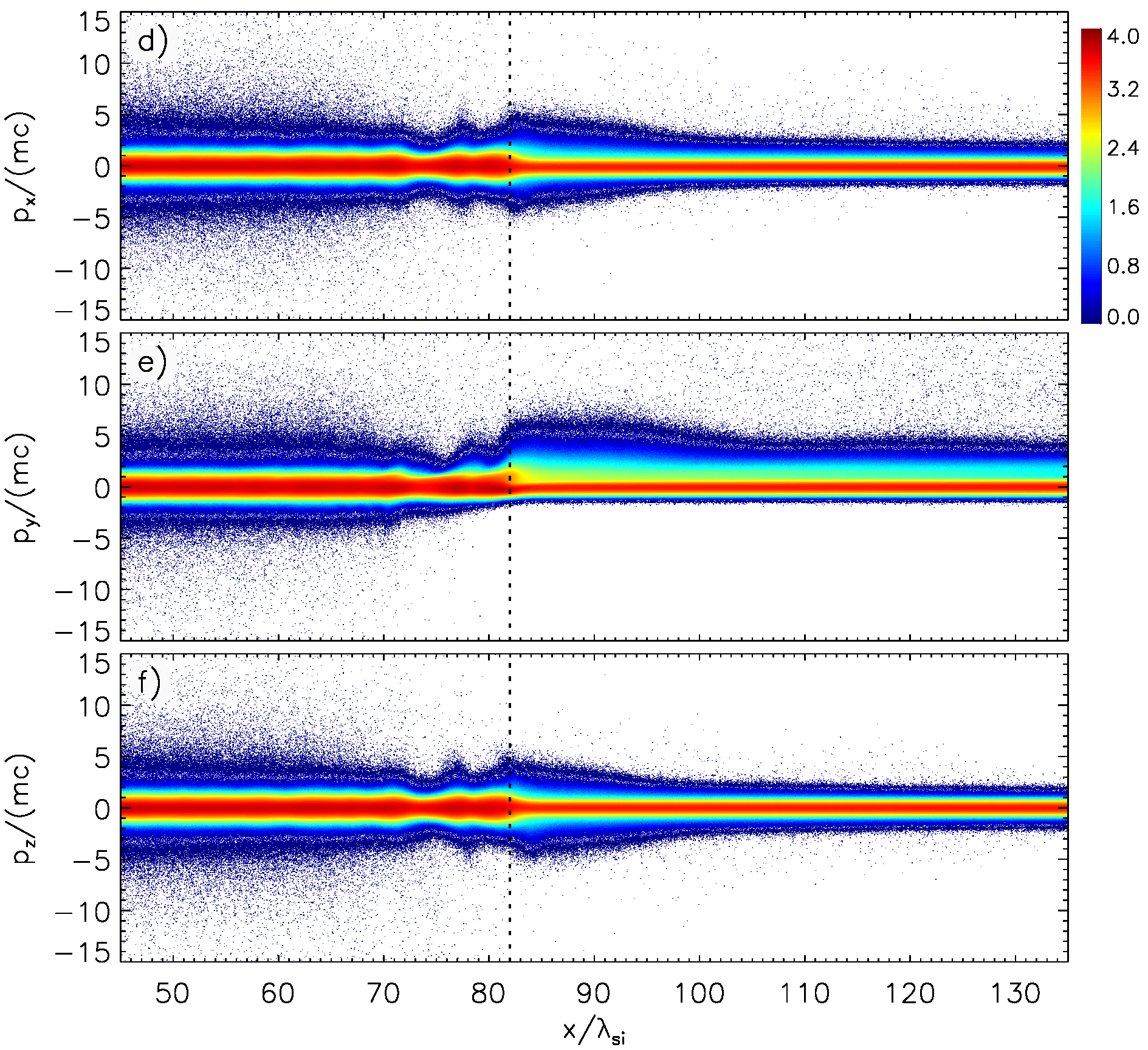}
\caption{Electron phase-space distributions at the \textit{laminar} (left, $\Oi t = 18$) and \textit{rippled} (right, $\Oi t = 36$) stage of shock evolution. From top to bottom: $p_{x}/mc$ (\textit{a}, \textit{d}), $p_{y}/mc$ (\textit{b}, \textit{e}) and $p_{z}/mc$ (\textit{c}, \textit{f}), all averaged in $y$-direction. The vertical dotted lines at $x \approx 43.5 \lsi$ and at $x \approx 82\, \lsi$ denote the shock location.}
\label{fig:phase}
\end{figure*}

In this section we present
the evolution of the shock structure, focusing on the role of the shock-front corrugations in the formation of multi-scale turbulence, that is of profound importance for electron acceleration. In our numerical experiment shock rippling appears at time $\Oi t \approx 25$ and is well developed by $\Oi t \approx 36$. 
Figures~\ref{fig:maps} and \ref{fig:phase} compare the shock structure at times $\Oi t = 18$ (\textit{left}) and $\Oi t = 36$ (\textit{right}), representing the early \textit{laminar} and the later \textit{rippled} stage, respectively. Maps of the electron density and the $B_z$ and $E_x$ field components shown in Figure~\ref{fig:maps} provide illustration of the waves present in the shock transition. Overplotted with a solid black line in the bottom panels c) and f) is the normalized, $y$-averaged value of the electric potential energy, 
\begin{equation}
  \phi = -\frac{e}{m_{\mathrm{e}}c^{2}} 
\int_\infty^x \langle  E_x(x') \rangle\,dx'\, .
\label{eqphi}
\end{equation}

Figure~\ref{fig:phase} shows the corresponding electron phase-space distributions. At time $\Oi t = 18$ the shock is located at $x\approx 43.5\, \lsi$, and it moves to $x \approx 82\, \lsi$ by $\Oi t = 36$. By then the shock has already assumed its quasi-stationary form and propagates with velocity $v_{\mathrm{sh}} \simeq 0.05\,c$ (or $v_{\mathrm{sh}}^{\mathrm{up}} \simeq 0.15\,c$ measured in the upstream plasma rest frame). The shock shows an overshoot-undershoot structure that is typical for quasi-perpendicular shocks and is caused by the ion dynamics at the shock. The \emph{first overshoot} has two characteristic peaks, the forward one located at $x \approx 42\,\lsi$ at $\Oi t = 18$. This structure is followed by the \emph{undershoot} at $x\approx 34\,\lsi$ and the \emph{second overshoot} at $x\approx 31\,\lsi$.
The density compression ratio reaches $r_{\mathrm{sh}} \approx 3.5$ at the overshoots and relaxes to $r = 3$ further downstream, in agreement with the Rankine-Hugoniot conditions for an unmagnetized shock with Mach number $\ms=3$ and $\Gamma=5/3$.
The magnetic-field compression in the overshoot, $b_{\mathrm{sh}}=B/B_0 \approx 3.5$, is the same as that of density. 

At time $\Oi t = 36$ the shock ripples at the first overshoot have the wavelength $\lambda_{\mathrm{rippl}}\approx 16\,\lsi$, clearly visible as two maxima in the density map in Figure~\ref{fig:maps}d. The second overshoot is also corrugated. Ripples in this region emerge much earlier (compare Fig.~\ref{fig:maps}a) and have shorter wavelengths than those at the first overshoot, but with time the wavelength increases. 

The ripples significantly modify the shock transition. To be noted from Figure~\ref{fig:phase} are asymmetric wings in the $x-p_x$ and $x-p_y$ electron phase-space distributions that are present upstream of the shock at all times. They are formed by electrons reflected from the shock in SDA interactions. They move along the upstream magnetic field that has a dominant component along $y$-axis, hence the large asymmetry in $x-p_y$ phase-space. At the rippled shock (right panels) the wings are wider within about $20\lsi$ upstream of the shock than they are farther away. Far upstream the excess of electrons with large positive $p_x$ and $p_y$ is comparable to that at the laminar shock, probably because these electrons have been reflected at the shock during its laminar phase. We conclude that the ripples enhance the rate and momenta of reflected electrons, the latter to $p_{\mathrm{e}}/(m_{\mathrm{e}}c) \approx 7$.

To understand the increased electron reflection at the rippled shock, we analyze the initial pitch angle an inbound electron must have to be reflected in one SDA cycle. In the de Hoffman-Teller (HT) frame \citep{dHT-50}, the initial pitch angle must satisfy
\begin{equation}
{\alpha_\mathrm{i}^\mathrm{HT}\equiv\cot\frac{v_{\mathrm{i},\perp}}{v_{\mathrm{i},\parallel}}\geq\sin^{-1}\sqrt{\frac{\left[\gamma_\mathrm{i}^\mathrm{HT}+\Delta\phi^\mathrm{HT}\right]^2-1}{b^\mathrm{HT}\,\left([\gamma_\mathrm{i}^\mathrm{HT}]^2-1\right)}} }~,
\end{equation}
with velocity components measured with respect to the background magnetic field, $\bm{B}_0^\mathrm{HT}$, the normalized cross-shock potential jump, $\Delta\phi^\mathrm{HT}\equiv e[\phi^\mathrm{HT}-\phi_0^\mathrm{HT}]/m_\mathrm{e}c^2$, and the magnetic-field compression 
$b^\mathrm{HT}\equiv B^\mathrm{HT}/B_0^\mathrm{HT}$, where both $B^\mathrm{HT}$ and $\phi^\mathrm{HT}$ are measured at the overshoot.
It follows that with larger compression or smaller potential drop the minimum pitch angle for SDA decreases, allowing more incoming electrons to experience SDA. To be noted in Figures~\ref{fig:maps}c and ~\ref{fig:maps}f is that at the rippled shock $\Delta \phi$ is smaller by a factor of $1.3$ compared to the laminar stage.
Even if we allow for shifts in $x$-direction on account of shock corrugations, $\Delta \phi$ is still smaller by the factor $1.2$.
The average magnetic compression at the overshoot of the rippled shock is smaller than that at the laminar one by a factor of $1.17$, $\langle b_{\mathrm{sh}}\rangle\approx 3$.
These scalings hold in the HT frame, since both the cross-shock potential and the magnetic compression have similar values in HT and the simulation frame. The effects of magnetic compression thus largely compensate the average drop in the electric potential energy. However, the compression varies along the shock surface from $b_{\mathrm{sh,min}} \approx 2.8$ to $b_{\mathrm{sh,max}} \approx 3.6$. In the regions of stronger compressions the electron reflection can therefore be enhanced (see also below).

Comparison of the $B_z$ maps in Figures~\ref{fig:maps}b and~\ref{fig:maps}e reveals that in the rippled phase the magnetic waves upstream of the shock are significantly stronger. As we discuss in detail in Section~\ref{subsec:waves}, these waves are the oblique modes of the EFI, driven by the effective electron temperature anisotropy that is provided by SDA-reflected electrons streaming along the magnetic field. The low amplitude of these waves at the laminar shock is in line with recent finding that few electrons are reflected, and the resulting weak temperature anisotropy provides inefficient wave generation \citep{Guo-14b}, if 
$v_{\mathrm{t}} \gtrsim
v_{\mathrm{th,e}}$, where $v_{\mathrm{t}} = v_{\mathrm{sh}}^{\mathrm{up}}/\cos \theta_{\mathrm{Bn}}$ is de Hoffman--Teller velocity and $v_{\mathrm{th,e}}$ the thermal speed of upstream electrons.
 
In our simulation we have $v_{\mathrm{t}} \approx 1.5\,v_{\mathrm{th,e}}$, and so electron reflection should be moderately suppressed. The limiting obliquity angle, ${\theta_{\rm limit}}=\arccos\left(\ms\sqrt{\Gamma m_\mathrm{e}/m_\mathrm{i}}\right)\simeq 67^{\circ}$, is not far from the magnetic obliquity in the simulation, and so small changes of the local obliquity angle caused by the shock ripples and combined with increased magnetic field compression can provide localized efficient electron reflection and EFI driving.
The observed shock corrugations change the local obliquity by up to $10^{\circ}$. Since these corrugations are asymmetric, about $2/3$ of the shock surface has an obliquity $\theta_{\mathrm{Bn}} \lesssim 75^\circ$, and the conditions might be favorable for electron reflection.
The modulation of the EFI wave amplitude along the shock, that is evident in Figure~\ref{fig:maps}e, is consistent with this expectation.

The electron and ion density distributions are generally well correlated, except for the small-scale waves in the upstream region beyond $x \approx 60\, \lsi$ that have associated electric-field fluctuations (weak in Fig.~\ref{fig:maps}c). These \textit{electrostatic} waves propagate upwards approximately along the large-scale magnetic field. We confirmed that they are Langmuir waves \citep{T-L-29} generated via the electron bump-on-tail instability \citep{Sarkar-15}, that here is driven by reflected electrons.
There is no evidence of any influence on electron acceleration, the main subject of this article, and so we do not discuss them in detail.

\begin{figure}
\centering
\includegraphics[width=0.99 \linewidth, clip]{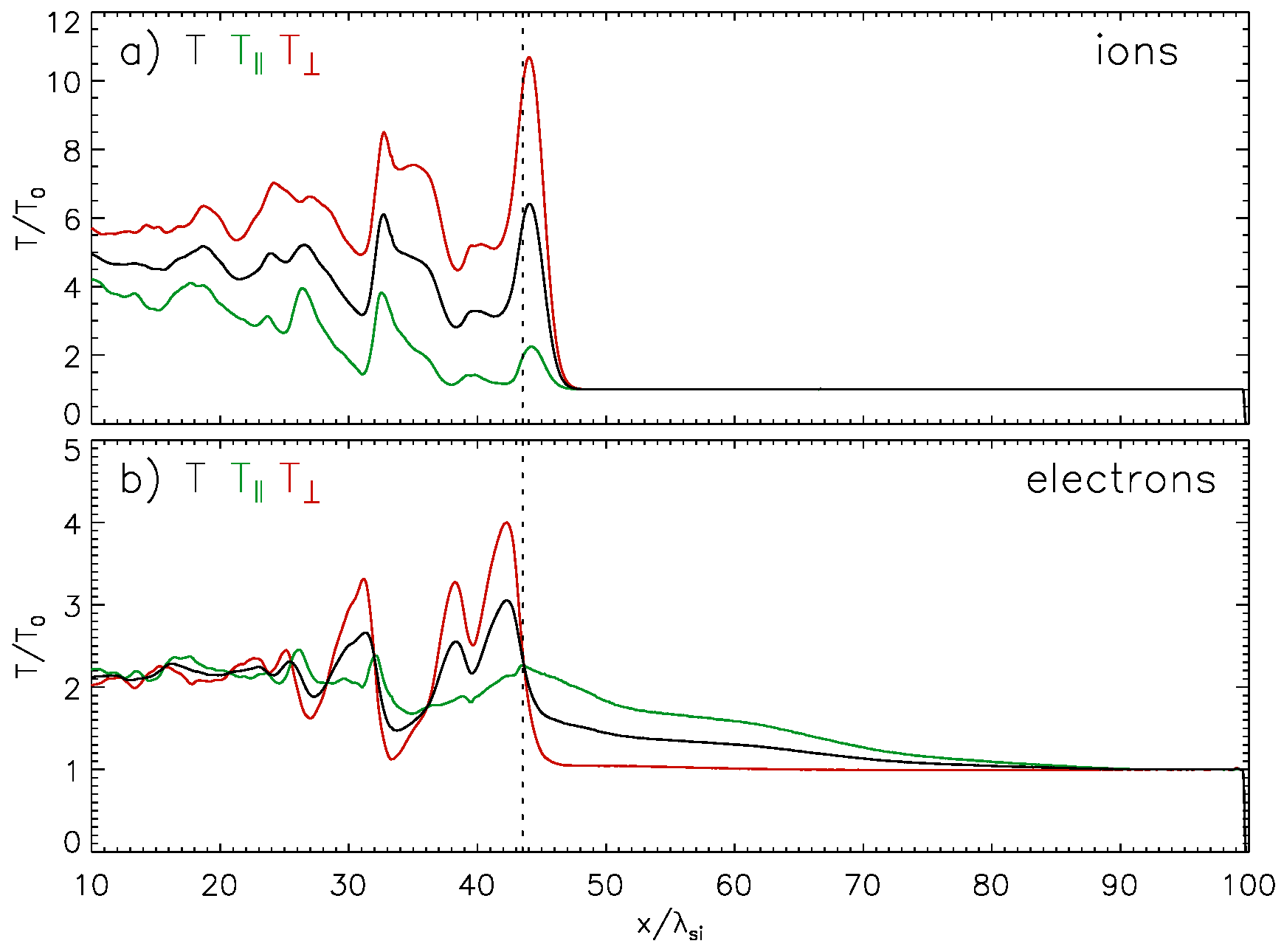}
\caption{Normalized ion (\textit{a}) and electron (\textit{b}) temperature profiles at $\Oi t = 18$, as well as their components parallel and perpendicular to the local magnetic field. The vertical dotted line in each panel marks the shock location. 
}
\label{fig:temp}
\end{figure}

\subsection{Wave Turbulence}
\label{subsec:waves}

The maps of particle density and electromagnetic field amplitudes shown in Figure~\ref{fig:maps} reveal various wave modes in a wide wavevector range. As we demonstrate below, these wave modes are driven by temperature anisotropy of either ions or electrons at the shock transition.

\subsubsection{Temperature Anisotropy}
\label{subsec:T-ani}

Figure~\ref{fig:temp} shows profiles of ion and electron temperatures at the early, laminar shock at $\Oi t = 18$.
We consider the temperature components parallel, $T_{\parallel}$, and perpendicular, $T_{\perp}$, to the \emph{local} magnetic field, normalized so that $T_{\parallel} = T_{\perp} = \rev{T_0}$ far upstream of the shock.

One can see a strong anisotropy in the ion temperature, $T_{\mathrm{i}\, \perp}/T_{\mathrm{i}\, \parallel}\gg 1$, at the shock ramp and overshoot (Fig.~\ref{fig:temp}a). It is generated by ions reflected off the shock that gyrate in the upstream magnetic field and gain energy by drifting along the motional electric field, $E_{0z}\bm{\hat{z}}$. 
This energy gain enables the ions to overcome the cross-shock potential drop and be advected downstream upon a single reflection. The temperature anisotropy is therefore confined to within one ion gyroradius from the shock, $r_\mathrm{gi}\lesssim 5\,\lsi$. The temperature anisotropy persists downstream of the shock and its amplitude decreases with distance from the shock, as the ion distribution isotropizes through scattering off turbulence.

As discussed in Section~\ref{subsec:structure}, the electron temperature anisotropy upstream of the shock,
$T_{\mathrm{e}\, \parallel} > T_{\mathrm{e}\, \perp}$ (Fig.~\ref{fig:temp}b), arises because of the presence of reflected electrons streaming nearly parallel to the magnetic field.
Downstream of the shock the temperature anisotropy shows the opposite trend, $T_{\mathrm{e}\, \perp} > T_{\mathrm{e}\, \parallel}$, most prominently at the double-peaked first overshoot and at the second overshoot. Electron gyration is fast, and plasma heating in these regions is mainly due to adiabatic compression, and so the \textit{local} conservation of the magnetic moment (the 1-st adiabatic invariant), $\mu = p_{\mathrm{\perp}}^2/(2mB)$, results in dominant growth of the perpendicular momentum. Correspondingly, only in the undershoot, and also the second and third undershoots at $x/\lsi\approx 27$ and 18, respectively, we again have $T_{\mathrm{e}\, \parallel}\gtrsim T_{\mathrm{e}\, \perp}$. One can see in Figure~\ref{fig:phase}a-c, that in these regions there are populations of electrons that were reflected from the second, third, and the fourth overshoot. 

\begin{figure}
\centering
\includegraphics[width=0.8 \linewidth]{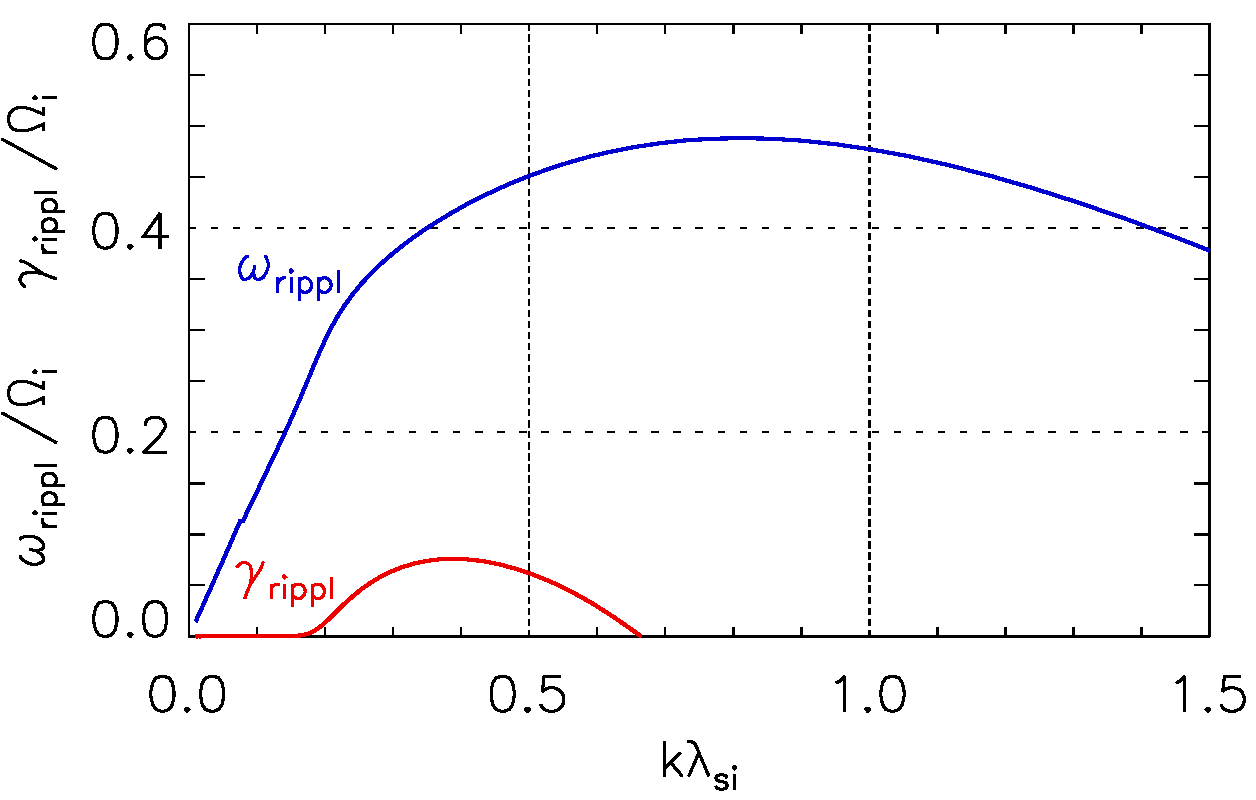}
\caption{Results of the linear dispersion analysis for the rippling modes: (blue) frequency, $\omega_{\mathrm{rippl}} (k)$, and (red) growth rate, $\gamma_{\mathrm{rippl}} (k)$.
}
\label{fig:ripple}
\end{figure}

\subsubsection{Ripple Mode}

Ion temperature anisotropy of direction $T_{\mathrm{i}\, \perp} > T_{\mathrm{i}\, \parallel}$ at the shock and downstream should trigger the AIC instability \citep{Winske-88, McKean-95, Lowe-03}, which is responsible for the emergence of the shock ripples. 
To estimate the expected properties of the ripple mode we have performed a linear dispersion analysis. 
This analysis assumes that the ion distribution is represented by two populations: isotropic transmitted ions and anisotropic reflected ions. The fraction of the reflected ions was estimated as $N_r/N_i = 0.25$. The temperature anisotropy, $T_{\mathrm{i}\, \perp}/T_{\mathrm{i}\, \parallel} = 4.7$, is due to the reflected ion component, which is assumed to have bi-Maxwellian distribution function. 
The results are shown in Figure~\ref{fig:ripple}, in which we plot the real frequency (blue line) and the growth rate (red line) as functions of wavevector component, $k_\parallel$, parallel to the magnetic field lines. The fastest growth occurs at $k_\parallel c/\opi =  0.38$, corresponding to wavelength of $16.5\, \lsi$, which is in good agreement with the observed wavelength, $\lambda_{\mathrm{rippl}} = 16\lsi$. 
The latter is the nearest wavelength allowed by the simulation grid. The broadband character of the AIC suggests a negligible influence of the wavelength limitation on the growth rate. The peak growth rate, $\gamma_{\mathrm{max}} = 0.076\, \Oi$, corresponds to two exponential growth cycles at time $\Oi t \approx 25$, at which the rippling modes appear in the simulation. In the simulation frame the rippling structure moves downwards along the shock surface with velocity $v_{\mathrm{rippl}} \approx 0.06\, c$, that is close to the Alfv\'{e}n velocity in the overshoot. The observed ripple waves can thus be firmly identified with AIC modes.

The gradual decrease in the ion temperature anisotropy in the region beyond the first overshoot may be understood in terms of relaxation via the AIC instability.
The downstream electromagnetic structure suggests that also 
the mirror instabilities may operate there.

\begin{figure}
\centering
\includegraphics[width=0.99\linewidth, clip]{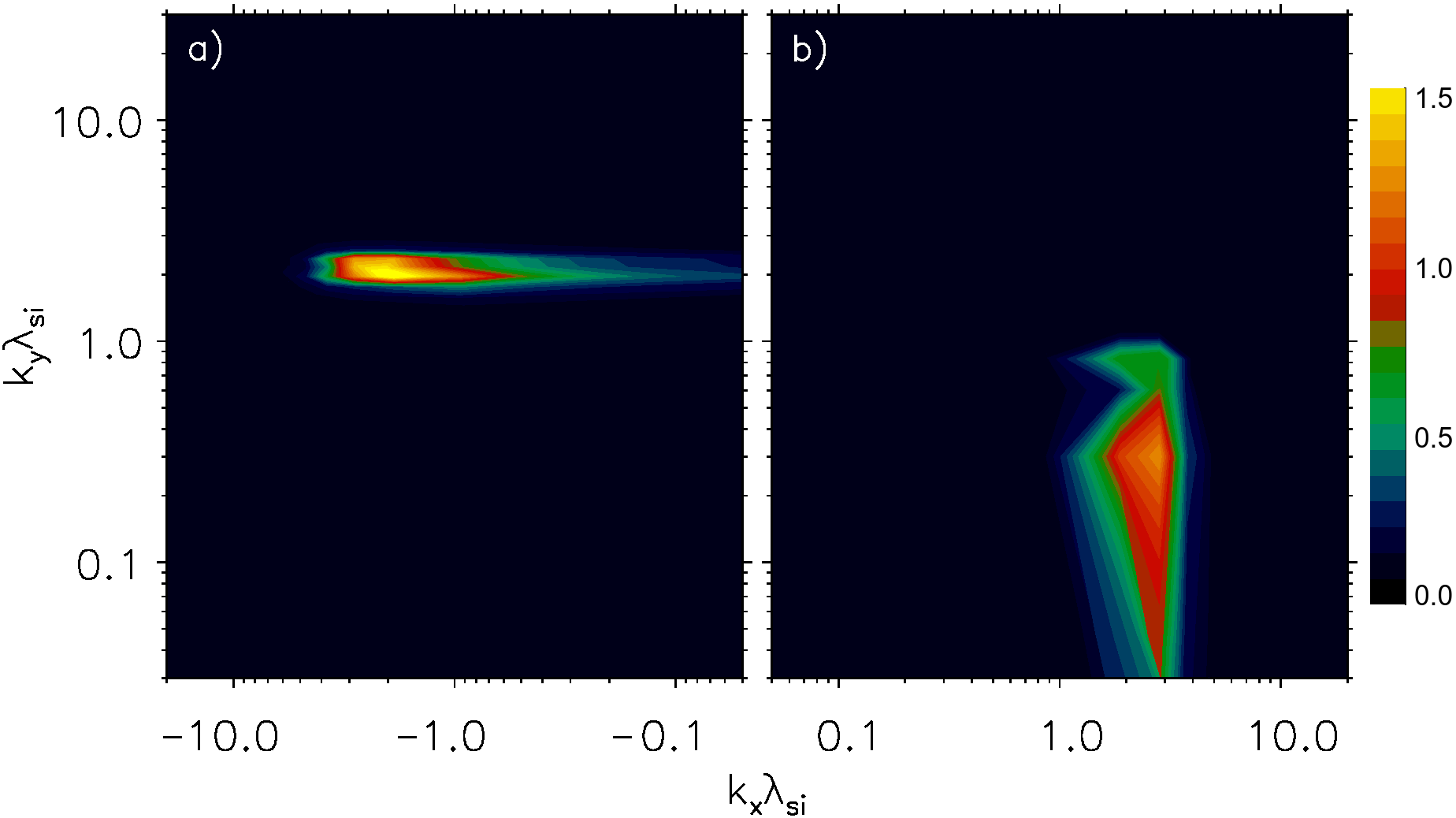}
\caption{Fourier power spectra for the magnetic waves ahead of the rippled shock at $\Oi t = 36$, calculated for the region $85\, \lsi < x < 95\, \lsi$ in Fig.~\ref{fig:maps}e.
}
\label{fig:four_B}
\end{figure}

\subsubsection{EFI Waves}

We noted before that the upstream magnetic waves, that are visible mainly in the $B_{z}$ component ($|\delta B_z|\gg |\delta B_x|$, $|\delta B_y|$) and amplified at the emergence of the shock rippling (Fig.~\ref{fig:maps}b and ~\ref{fig:maps}e), are associated with EFI triggered by SDA-reflected electrons. They are in fact two oblique 
modes whose inclination is roughly symmetric with respect to the large-scale upstream magnetic field. Figure~\ref{fig:four_B} shows Fourier power spectra of these waves after their amplification at time $\Oi t = 36$. Panels a) and b) are calculated for the negative and positive wavevectors, $\bm{k}_x$.
The maximum wave power in waves in panel a) is at $(k_{x},k_{y})\, \lsi \approx (-2.4,2.0)$, which corresponds to $\lambda\approx 2.0\, \lsi$ and $\theta_{\mathrm{EFI}}\approx 66^{\circ}$, where $\theta_{\mathrm{EFI}}$ is the angle between the wavevector and the background magnetic field. The peak signal in panel b)
is at $(k_{x},k_{y})\, \lsi \approx (2.8,0.3)$, so that $\lambda\approx 2.2\, \lsi$ and $\theta_{\mathrm{EFI}} \approx 69^{\circ}$. Hence both wave components have approximately the same wavelength, $\lambda_{\mathrm{EFI}} \approx (2.1 \pm 0.1)\, \lsi$, and the same inclination angle with respect to the upstream magnetic field direction, $\theta_{\mathrm{EFI}} \approx (67.5 \pm 1.5)^{\circ}$. These characteristics are in agreement with upstream waves observed in PIC simulations of intracluster shocks by \citet{Guo-14a,Guo-14b} and \citet{Kang-19}, that demonstrated consistency with theoretical predictions for EFI driven by an electron temperature anisotropy \citep[e.g.,][]{Li-00, Camporeale-08}. The EFI wave properties are also in line with the electron beam (or heat flux) driven modes, that have recently been shown by \citet{Kim-20} to be more relevant for conditions at high-$\beta$ shocks. Both the temperature anisotropy and beam modes have similar properties that are difficult to distinguish in simulations. We could not detect propagation of the EFI waves in the upstream rest frame at a phase speed higher than the Alfv\'en speed, suggesting that their frequency is much smaller than their growth rate. This is consistent with the beam-driven EFI and in particular with the nonpropagating oblique EFI mode driven by temperature anisotropy.

\subsubsection{Whistler waves}

\begin{figure}
\centering
\includegraphics[width=0.98 \linewidth, clip]{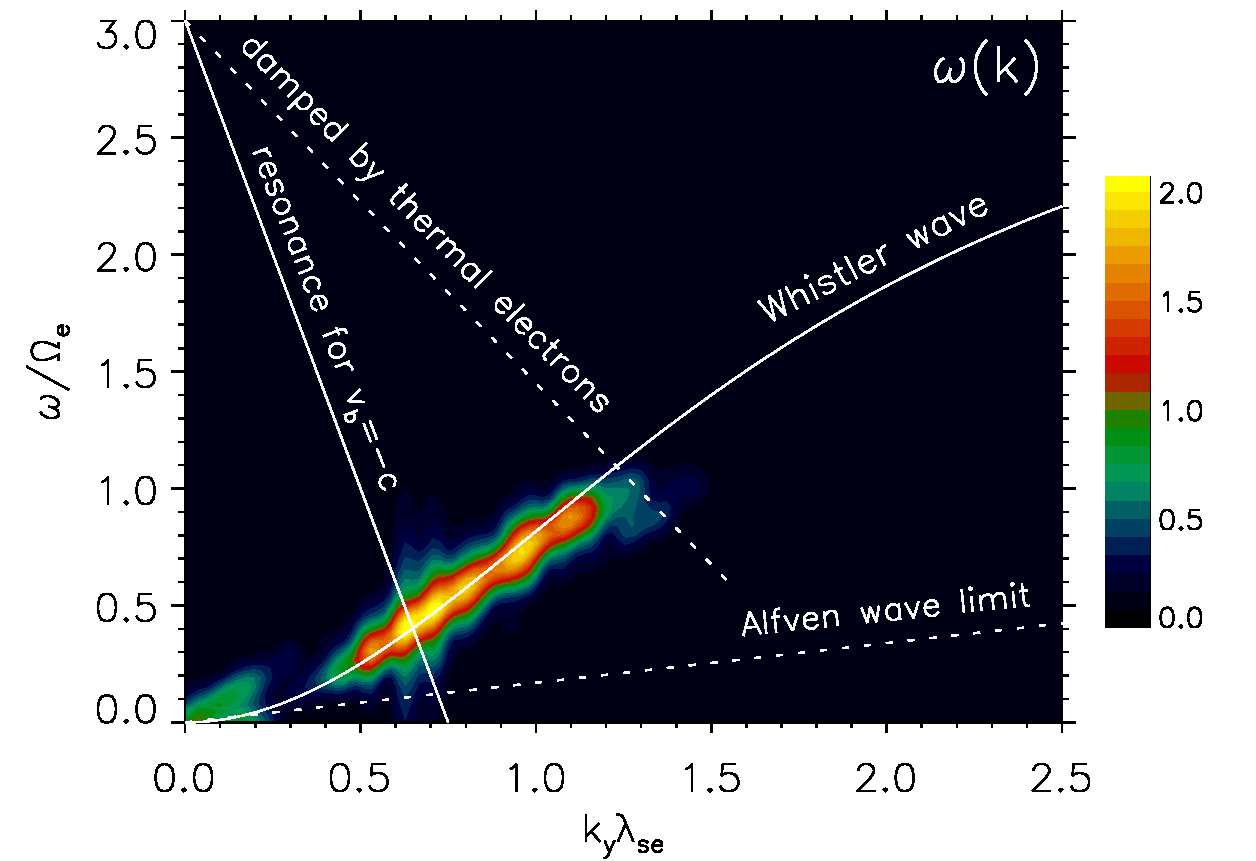}
\caption{Fourier spectrum of magnetic waves at, and co-moving with, the shock overshoot. 
The time interval is $1.63\, \Oi^{-1} = 163\, \Oe^{-1}$, starting from $t\Oi = 18$.} 
\label{fig:sp-time}
\end{figure}

The field maps in Figure~\ref{fig:maps} show small-scale waves at the first and the second overshoot. They have $B_{x}$ and $B_{z}$ field oscillations and propagate upwards along the shock-compressed magnetic field. They are prominent in regions with $T_{\mathrm{e}\, \perp} > T_{\mathrm{e}\, \parallel}$ (see Section~\ref{subsec:T-ani}) and most likely 
right-hand circularly-polarized whistlers excited by the \emph{electron} temperature anisotropy. 

Figure~\ref{fig:sp-time} presents a Fourier
analysis of the $B_{z}$ field oscillations at the first overshoot.
For convenience, the axes are scaled with the electron skin depth, $\lse = 0.1\, \lsi$, and the electron gyro-frequency, $\Oe = 100\, \Oi$. 
The starting time for the analysis is $t = 18\, \Oi^{-1} = 1800\, \Oe^{-1}$, at which the waves are located at
$x/\lsi\approx 42$  (see Fig.~\ref{fig:maps}b) and the shock is still laminar. For $\Delta t \approx 1.63\, \Oi^{-1} = 163\, \Oe^{-1}$ the waves are followed co-moving with the shock, yielding the $\omega-k$ power spectrum that is calculated in the local plasma rest frame.

The main signal at $k \lse \approx (0.5-1.1)$ and $\omega \approx (0.3-0.9)\,\Oe$ can be identified with whistler waves.
In the low-frequency limit, \ok{$\omega \ll \Oe^{\mathrm{loc}}$}, 
the dispersion relation of whistler waves may be written in simplified form \ok{\citep{Bashir-12}:
\begin{equation}
\omega \approx \frac{\Oe^{\mathrm{loc}}\, c^{2} k_{\parallel}^{2}}{(\ope^{\mathrm{loc}})^{2} + c^{2} k_{\parallel}^{2}}
\left[ 1 + \frac{\beta_{\parallel}}{2} \left( \frac{T_{\mathrm{e}\,\perp}}{T_{\mathrm{e}\,\parallel}} - 1 \right) \right] ,
\label{eq:whist}
\end{equation}}
where $k_{\parallel}$ is the field-aligned wavevector and $\ope^{\mathrm{loc}}$ and $\Oe^{\mathrm{loc}}$ are, respectively, the \textit{local} electron plasma- and gyro-frequencies, which must be calculated with the \textit{average} magnetic field in the overshoot region that is compressed by a factor of $\langle b_{\mathrm{sh}}\rangle \approx 3$, a bit less than the maximum compression in a laminar shock on account of averaging over a region of $\sim 3\, \lsi$ in thickness.
The expected $\omega(k)$ dependence calculated from Equation~\ref{eq:whist} with parameters measured in the simulation is shown as a white solid curve in Figure~\ref{fig:sp-time}. To be noted is the good agreement between the observed and theoretical behavior.


Whistlers may be excited, if a beam of electrons with sufficiently large anisotropy $T_{\mathrm{e}\,\perp}/T_{\mathrm{e}\,\parallel} > 1$ (or a loss-cone) satisfies the cyclotron resonance condition:
\begin{equation}
\omega = \Oe + kv_{\mathrm{b}}~,
\label{eq:cycl}
\end{equation}
where $v_{\mathrm{b}}$ denotes the electron beam velocity parallel to the magnetic field \citep{Tokar-84, Amano-10}. 
The resonance condition for $v_{\mathrm{b}} = -c$ is shown in Figure~\ref{fig:sp-time} for reference with the solid straight line for parameters measured at the overshoot. Note that the negative beam velocity indicates that the electron beam propagates opposite to the waves. The resonance with a non-relativistic electron beam should occur to the right of this line. On the other hand, the wave growth will be suppressed by cyclotron damping of thermal electrons, which occurs at $\omega \gtrsim \Oe - kv_{\mathrm{th,e}}$. Therefore, one expects to observe wave signals related to the instability between the two resonance conditions, which is in good agreement with the simulation results.

A close look at the electron velocity distribution function in this region finds that the bulk of upstream electrons are accelerated to $v_b \sim -0.4 c$ (i.e., in the opposite direction to the waves) probably by the cross-shock electrostatic potential and are heated also adiabatically in the perpendicular direction. Such drifting anisotropic electrons are likely to be the cause of the instability on the whistler-mode branch \citep{Tokar-84}.

\section{Evolution of electron spectra}
\label{sec:distr}

We noted in the Section~\ref{sec:evolution} that shock rippling affects the electron phase-space distribution (Fig.~\ref{fig:phase}). In this section we discuss in detail the energy spectra of electrons.

\begin{figure}
\centering
\includegraphics[width=0.98 \linewidth, clip]{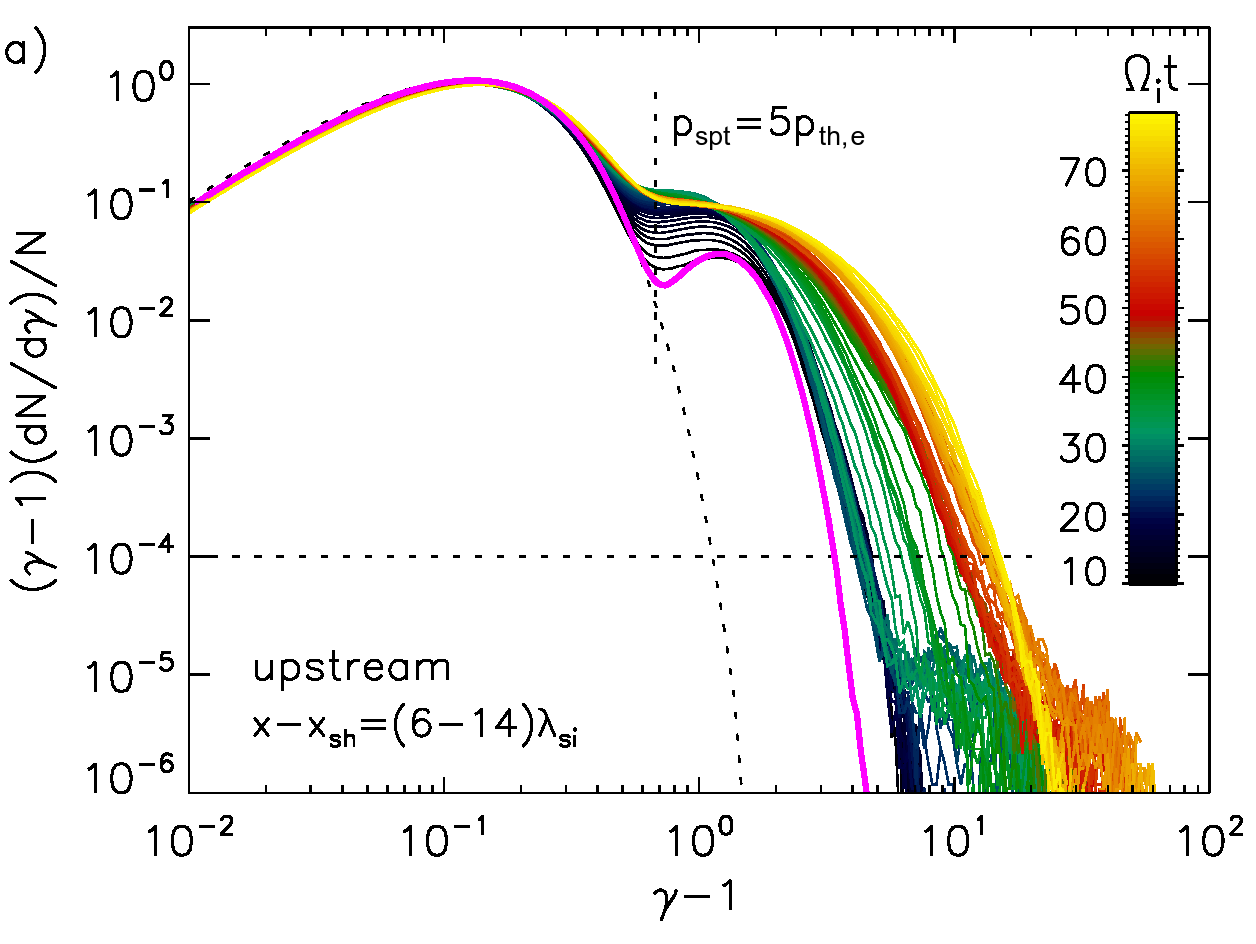}
\includegraphics[width=0.98 \linewidth, clip]{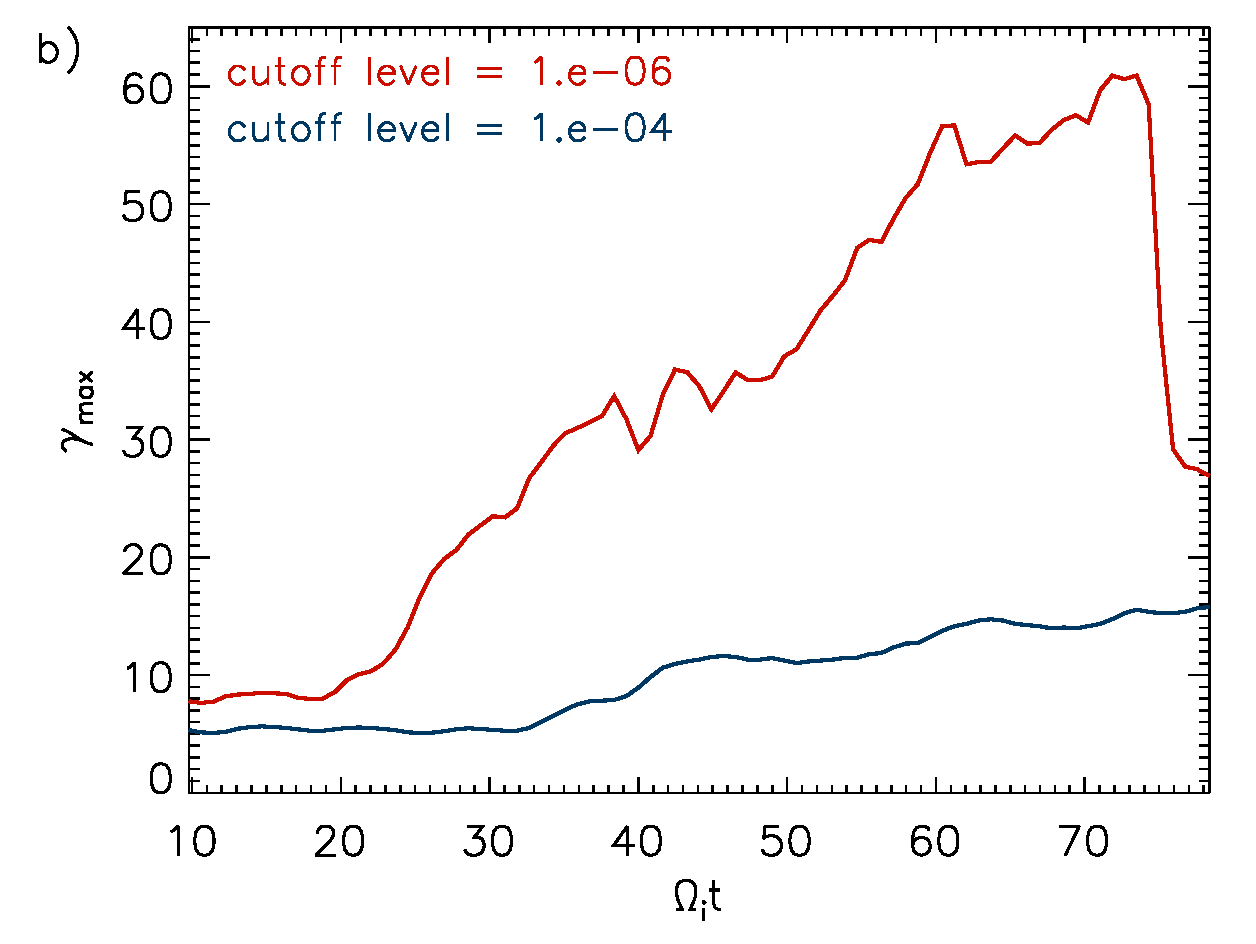}
\caption{(a): Energy spectra of upstream electrons at various times marked with color. The magenta curve displays the prediction of SDA theory. The dotted line is a Maxwellian fit to the  low-energy part of the spectra. 
(b): Evolution of the maximum Lorentz factor, $\gamma_{\mathrm{max}}$, for two cutoff levels.}
\label{fig:spectra}
\end{figure}

\subsection{Upstream Electron Spectra}
\label{subsec:distr-upstr}

Figure~\ref{fig:spectra}a shows the time evolution of the electron energy spectra in the region $ (6-14) \lsi$ \textit{upstream} of the shock. 
Figure~\ref{fig:spectra}b shows the evolution of the maximum Lorentz factor of electrons, $\gamma_{\rm max}$, for two cutoff levels of $(\gamma-1)(dN/d\gamma)/N$, \emph{blue} for $10^{-4}$ and \emph{red} for $10^{-6}$. The red curve traces the evolution of the most energetic electrons.

Supra-thermal electrons are produced already at the early laminar shock, $\Oi t \ll 25$, on account of SDA. We used the method of \citet{Guo-14a} to compute a synthetic spectrum of electrons accelerated in a single SDA cycle. It is shown as solid \emph{magenta} line in Figure~\ref{fig:spectra}. The match with the observed spectra is reasonably good, given that already at this phase processes other than SDA may energize electrons.

The energization rate, $d\gamma_{\rm max}/dt$, increases considerably upon
the appearance of rippling at $\Oi t \approx 25$ \rev{(red line in Fig.~\ref{fig:spectra}b)}, and a low-density population in particles with Lorentz factor of a few tens develops. 
\ok{The blue line in Figure~\ref{fig:spectra}b}
indicates that the bulk of the supra-thermal population commences a slow shift only after $\Oi t \approx 36$, when the shock ripples are fully developed.


At the end of the simulation these two spectral components merge into an extended non-thermal spectral tail that has a SDA-like shape, but extends to much higher energy than the standard SDA theory predicts. Due to inherent curvature the supra-thermal tail of the spectra cannot be fitted with a single power-law (compare \citet{Guo-14a,Kang-19}). The final spectrum is flat with slope $p\approx 1$ at $\gamma\gtrsim 2$ and continuously steepens with increasing energy.

The maximum Lorentz factor reaches $\gamma_{\mathrm{max}} \approx 60$, which is well above that needed for injection into DSA, $\gamma_{\mathrm{inj}} \approx 25$, typically estimated as few times $p_{\mathrm{th,i}}/m_\mathrm{e} c$ for relativistic electrons \citep{Kang-19}.
Taking $p_{\rm spt}=5p_{\mathrm{th,e}}$, we estimate the fraction of the supra-thermal electrons as
\begin{equation}
    \zeta=4\pi\int_{p_{\mathrm{spt}}}^{p_{\mathrm{max}}}\bigg\langle\frac{f(p)}{N} \bigg\rangle \, p^2 dp~,
\label{eq:fraq}
\end{equation}
where $\langle f(p)/N\rangle$ denotes the volume-average of the normalized distribution function. 
The final fraction of supra-thermal electrons 
reaches $\rev{\zeta\simeq } 5\%$, and about $40\%$ of the electron kinetic energy is carried by these energetic particles.
\rev{Note, that electrons with Lorentz factors $\gamma\approx 40-60$ disappear at the end of the simulation. This is due to boundary conditions, as discusses in Section~\ref{sec:accel}.}

\subsection{Downstream Electron Spectra}
\label{subsec:distr-dwnstr}

\begin{figure}
\centering
\includegraphics[width=0.98 \linewidth, clip]{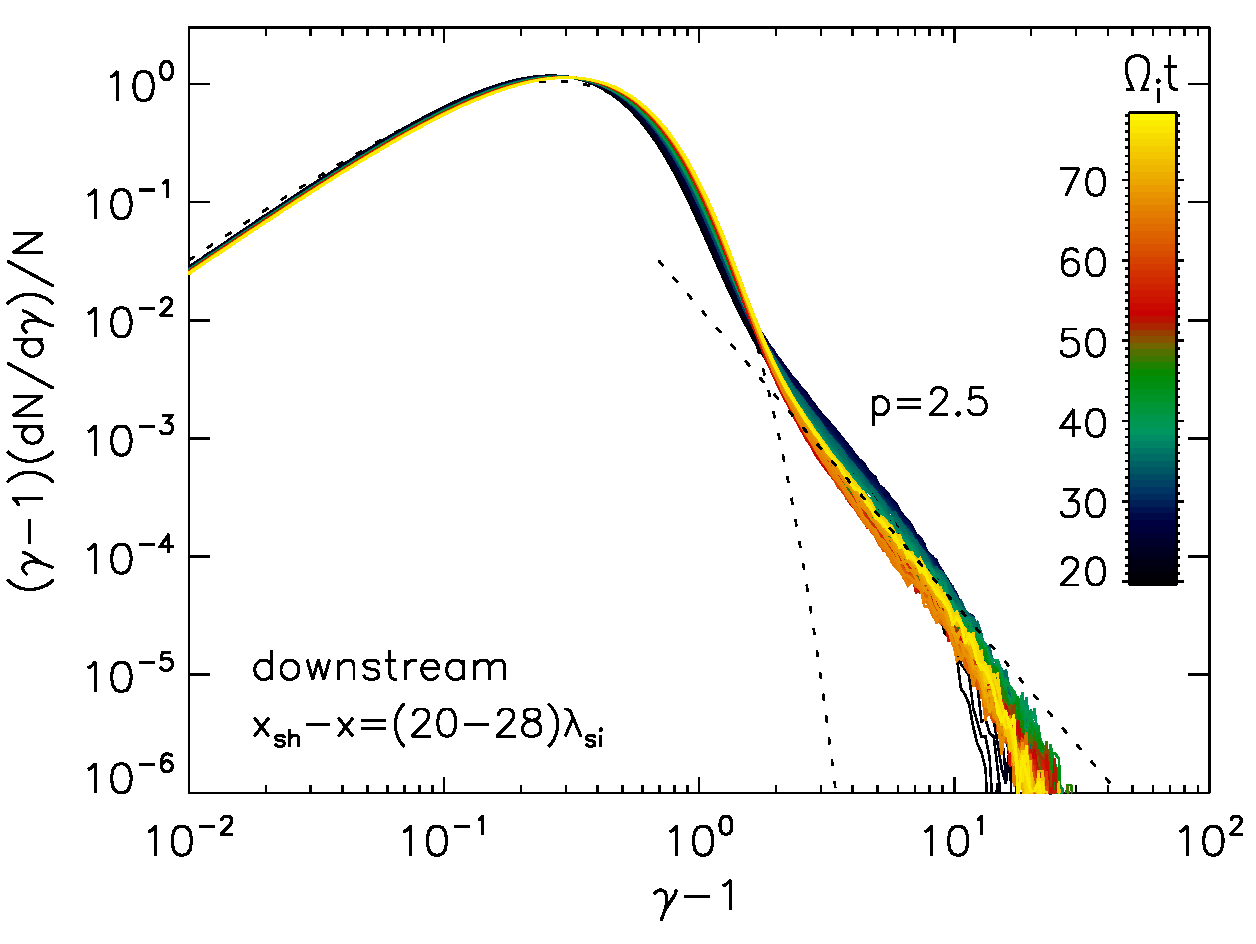}
\caption{Color-coded evolution of the downstream electron spectra. A Maxwellian fit to the low-energy part of the spectra is shown with a dotted line. The straight dotted line at high momenta denotes a power-law of slope \rev{$p = 2.5$}.
}
\label{fig:spectra-dwn}
\end{figure}

Figure~\ref{fig:spectra-dwn} shows the time evolution of the \textit{downstream} electron spectrum, measured in a region of thickness $\Delta x = 8\, \lsi$ that is located $20\, \lsi$ behind the shock. The spectra are typical for the entire downstream region, including the shock overshoot.  
An extended power-law tail is evident, which we demonstrate here for the first time. It appears already in the laminar shock phase and slowly evolves as time progresses. Later, with shock rippling in full operation, the spectral index settles at $p \approx \rev{2.5 \pm 0.1}$, and $\gamma_{\rm max}\lesssim 20$ stays approximately constant.
\rev{The spectral index is the same as one expects for test-particle DSA with the compression ratio $r=3$.}
The observed radio synchrotron spectra have \rev{also} a matching index, $\alpha = (p - 1)/2 \approx 0.75$ \citep[e.g.,][]{van-Weeren-10}, but would require much larger Lorentz factors, $\gamma\gg \gamma_{\rm max}$, than can be established in the short time that we simulate \rev{and using a finite-size simulation box}. The non-thermal tail contains about $0.12\%$ of particles and roughly $ 1\%$ of the electron energy.

\section{Electron acceleration}
\label{sec:accel}

\begin{figure}
\centering
\includegraphics[width=0.99 \linewidth, clip]{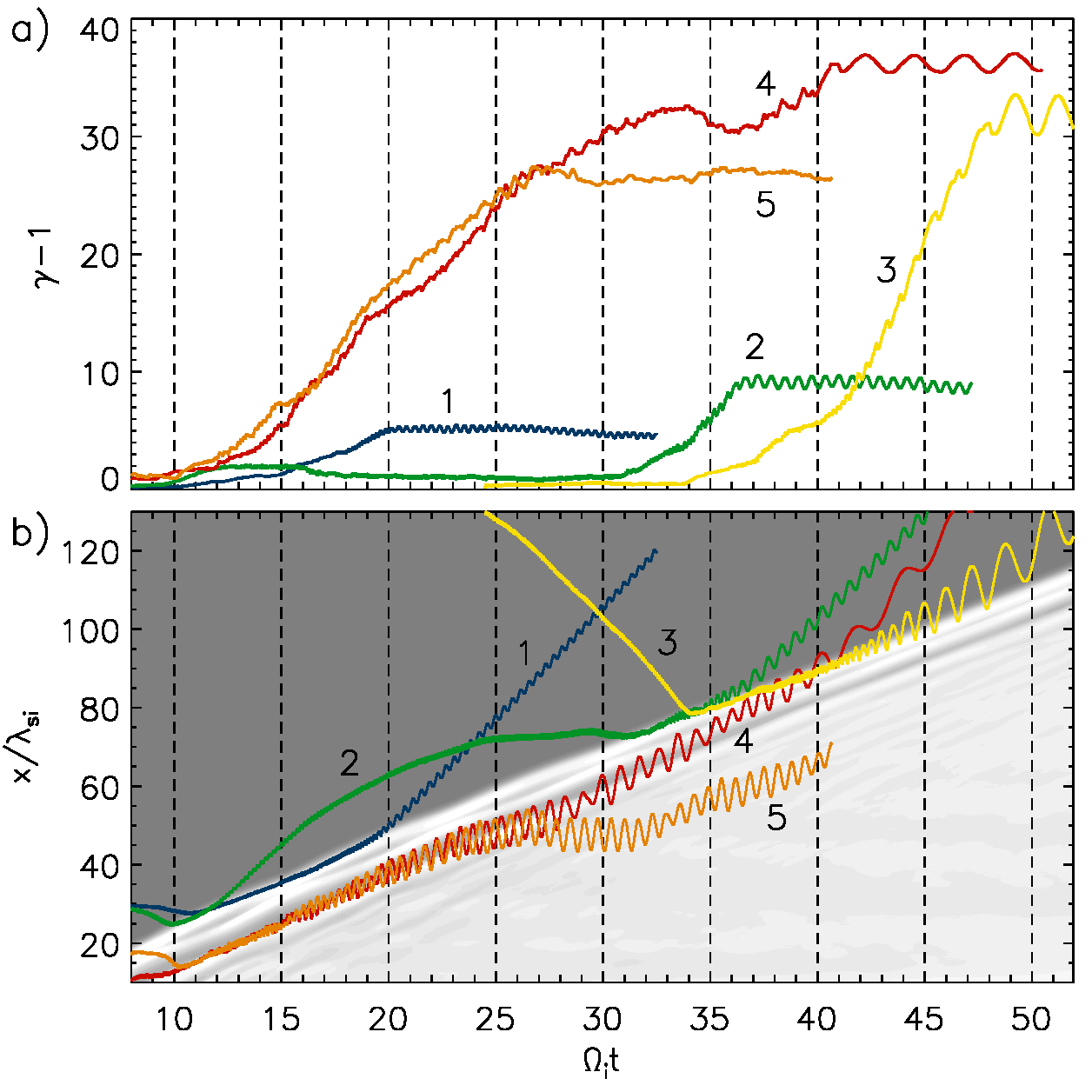}
\caption{Temporal evolution of particle energy (top) and particle trajectories on the background of the $y$-averaged density profile (bottom) for five typical electrons, color-coded and labeled as (1:blue), (2:green), (3:yellow), (4:red), and (5:orange).}
\label{fig:accel-multi}
\end{figure}

\begin{figure*}[t]
\centering
\includegraphics[width=0.49 \linewidth, clip]{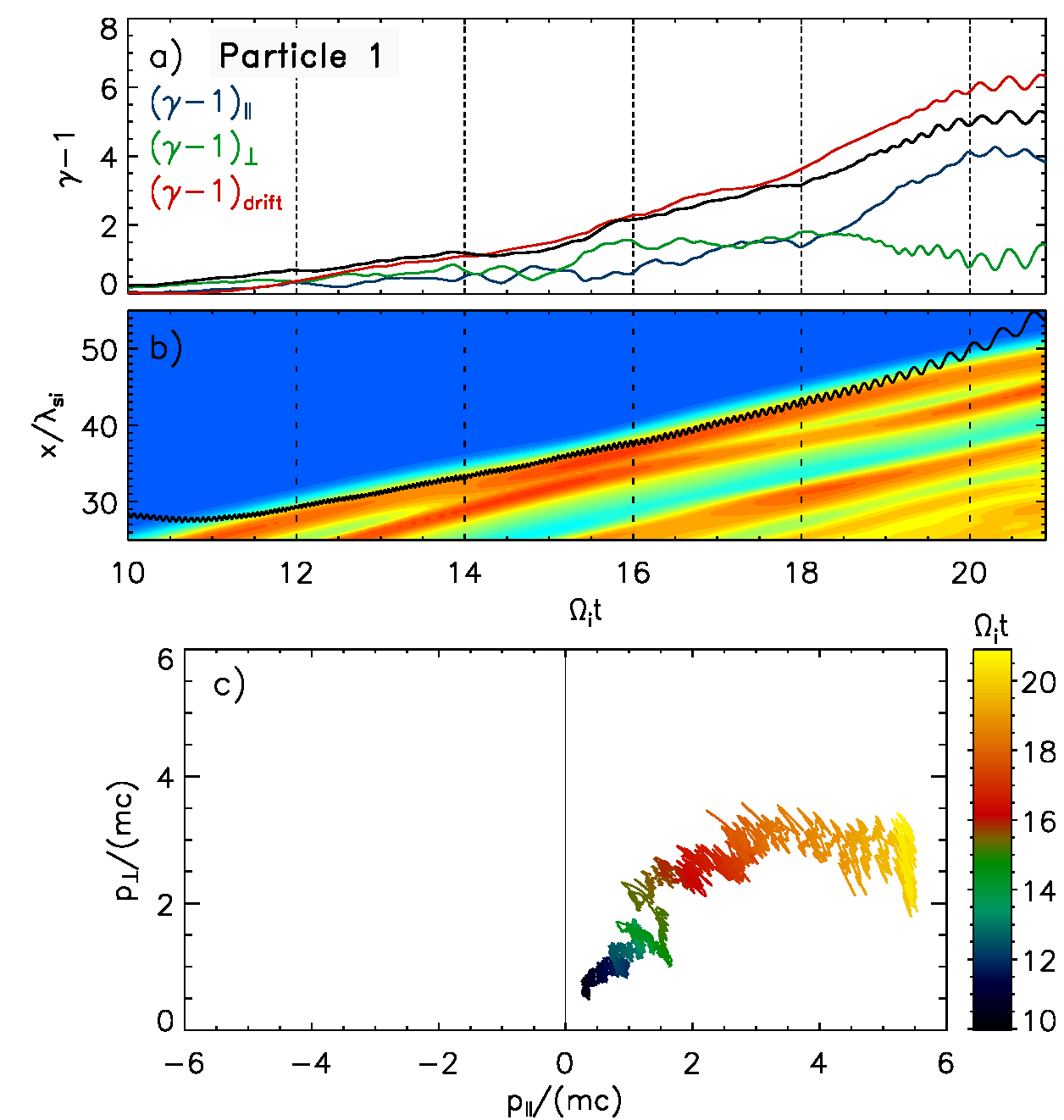}
\includegraphics[width=0.49 \linewidth, clip]{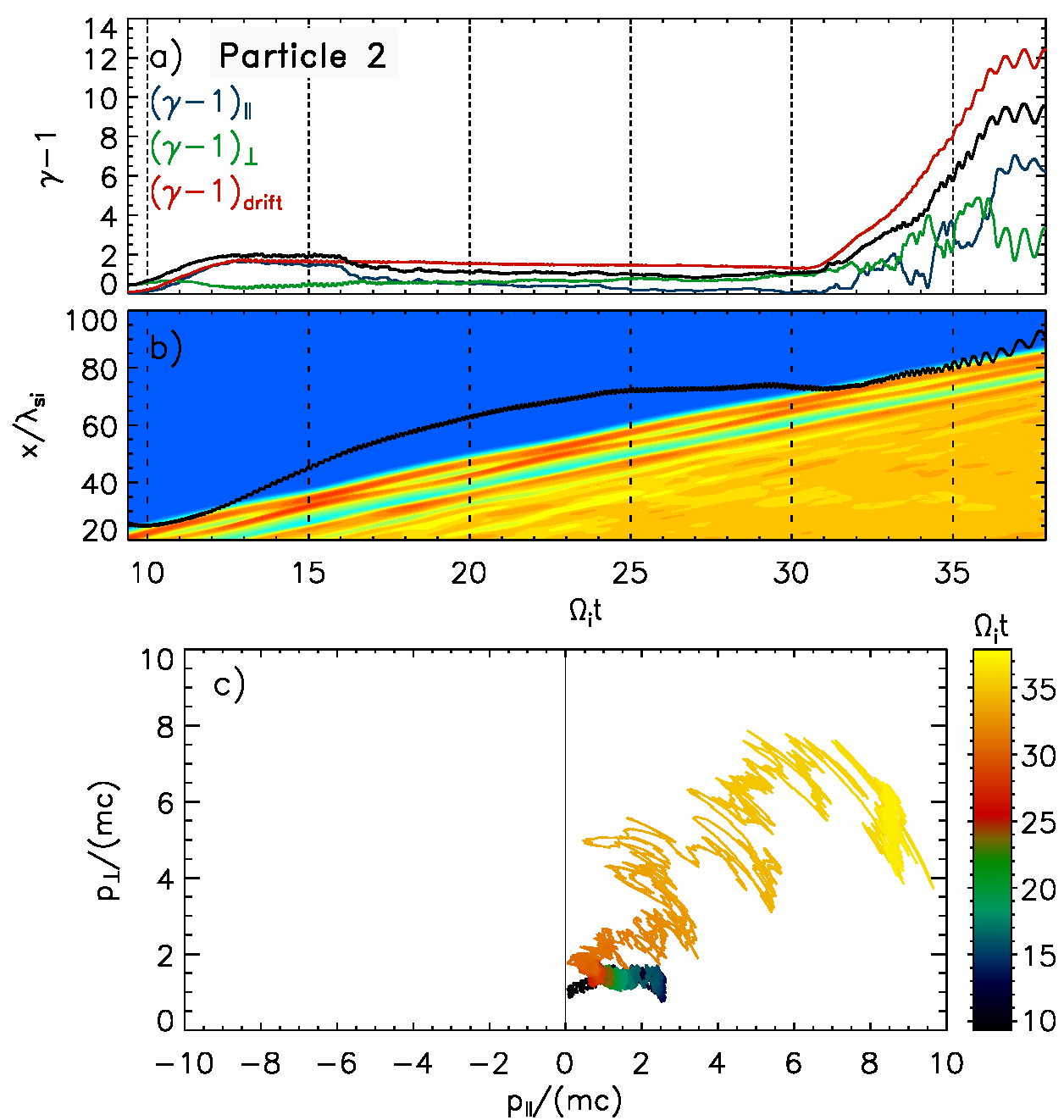}
\caption{Trajectories of particles 1 (blue) and 2 (green) shown in Fig.~\ref{fig:accel-multi}. Shown are: time evolution of energy  (\textit{black} line) and that expected for drift along the motional $E_{z}$-field (Eq.~\ref{eq:SDA}, \textit{red} line; {\it top} panels), the particle location relative to the shock in the $x$-direction overlaid on the $y$-averaged density profile ({\it middle} panels), and particle orbits in $p_{\parallel} - p_{\perp}$ momentum space with color-coded time-scale ({\it bottom} panels). All quantities are measured in the downstream rest frame.}
\label{fig:accel}
\end{figure*}

\begin{figure}[t]
\centering
\includegraphics[width=0.99 \linewidth, clip]{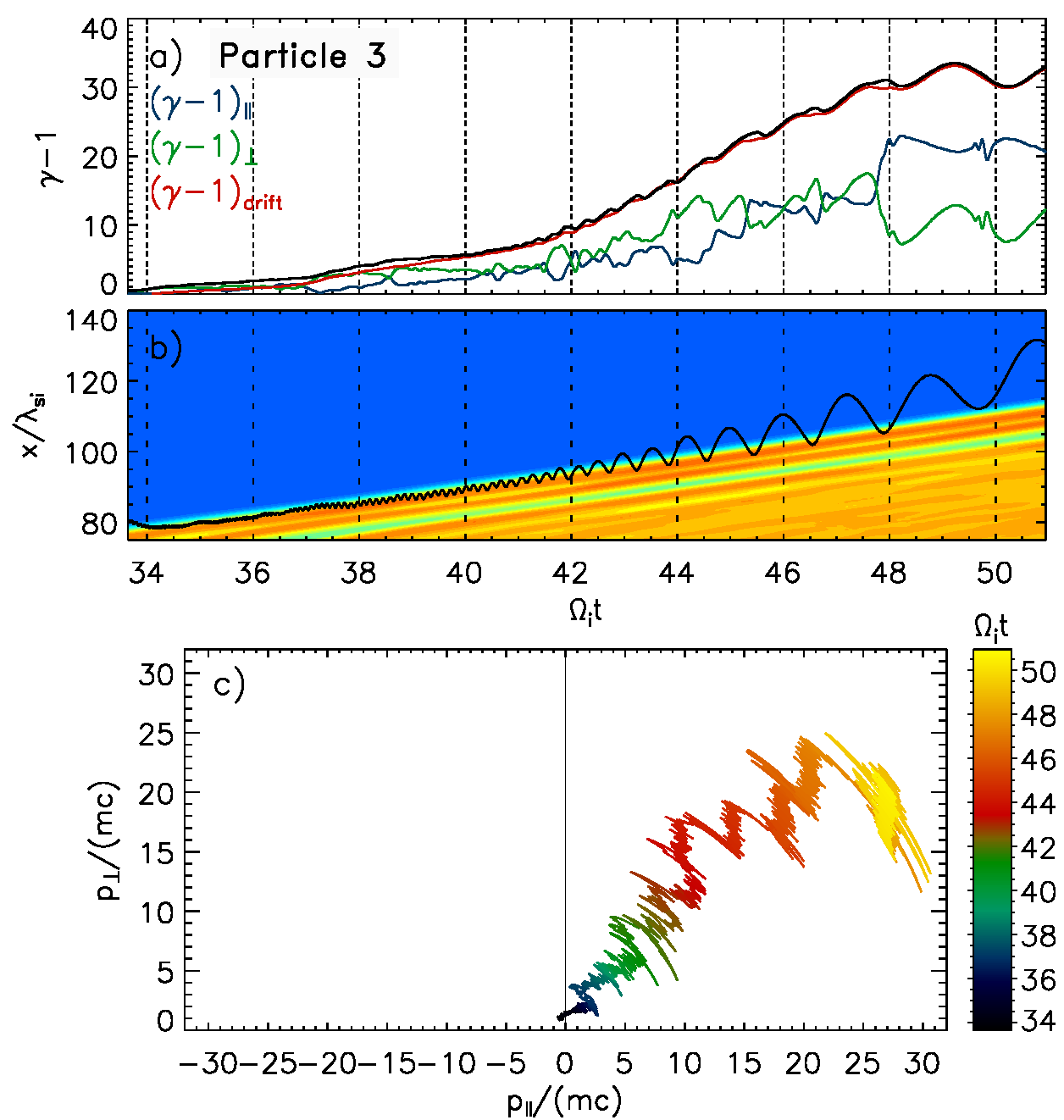}
\caption{Same as in Fig.~\ref{fig:accel} but for particle 3 shown in Fig.~\ref{fig:accel-multi} with yellow line.}
\label{fig:accel_3}
\end{figure}

\begin{figure*}[t]
\centering
\includegraphics[width=0.49 \linewidth, clip]{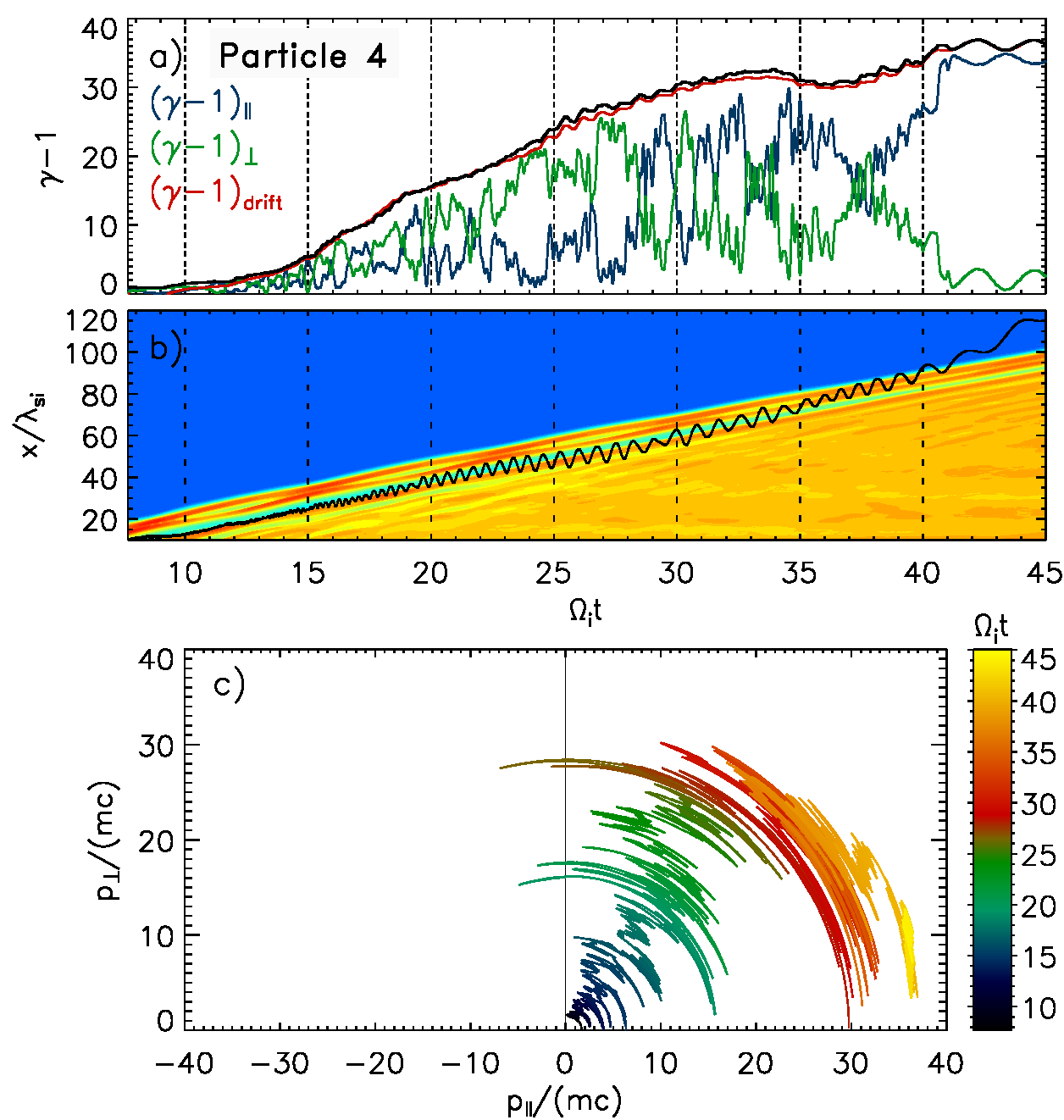}
\includegraphics[width=0.49 \linewidth, clip]{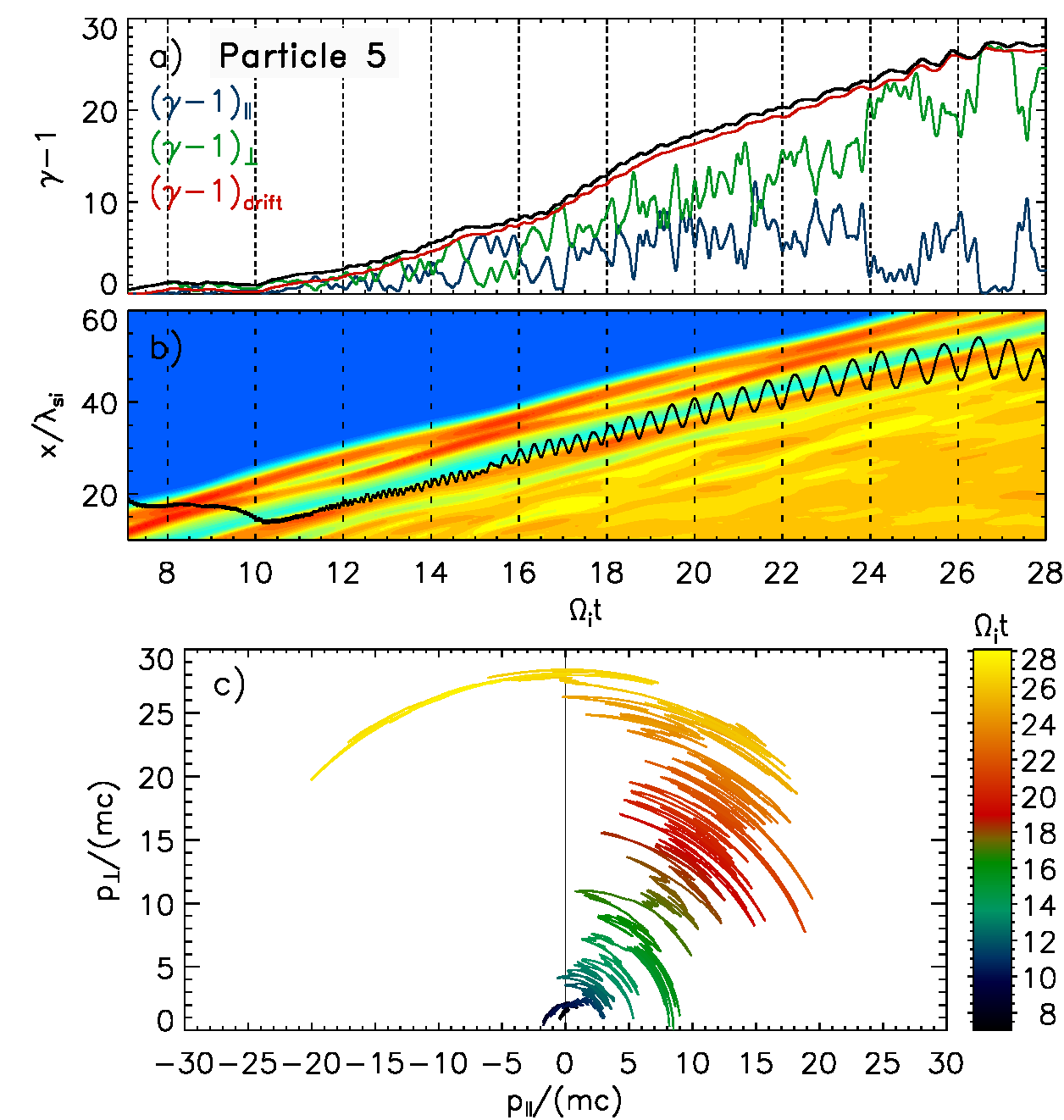}
\caption{Same as in Fig.~\ref{fig:accel} but for particles 4 and 5 shown in Fig.~\ref{fig:accel-multi} with red ({\it left}) and orange ({\it right}) lines.}
\label{fig:accel-dwn}
\end{figure*}

In this section we investigate the micro-physics of electron acceleration by tracing individual particles.
Figure~\ref{fig:accel-multi} displays trajectories of five typical accelerated electrons. Panel (a) shows the evolution of their kinetic energy, and panel (b) their location relative to the shock. 
The detailed evolution of the particle orbits is presented in Figures~\ref{fig:accel}-\ref{fig:accel-dwn}, in which we show the kinetic energy (\textit{black} line) in comparison with that expected from the drift anti-parallel to the motional electric field (\textit{red} line, Eq.~\ref{eq:SDA}). Parallel (\textit{blue}) and perpendicular (\textit{green}) parts of the kinetic energy are shown: $(\gamma - 1)_{\parallel} = (\gamma - 1) \cos^2 \alpha$ and $(\gamma - 1)_{\perp} = (\gamma - 1) \sin^2 \alpha$, where $\alpha$ is a pitch-angle.
Also shown in the middle and lower panels are the $x$-location relative to the shock and the trajectories in 
$p_{\parallel} - p_{\perp}$ momentum space, where $p_{\parallel}$ and $p_{\perp}$ are defined with respect to the mean magnetic field. All quantities are given in the simulation frame, in which the shock propagates slowly, implying that the shock and particle acceleration have similar properties as in the shock rest frame.

Most particles gain their energies during a single interaction with the shock. 
Double shock encounters (e.g., {\it particle~2}), resembling the process of multiple-cycle SDA, are rare due to inefficient scattering off upstream waves. Particle energization is achieved mostly through drift along the motional electric field, as in SDA (compare \textit{red} and \textit{black} curves in the \textit{top} panels of Figs. \ref{fig:accel}-\ref{fig:accel-dwn}).  
However, electrons typically do not undergo pure SDA, in which the acceleration time is of order $t_{\mathrm{SDA}} \sim \Oi^{-1}$ \citep{Krauss-Varban-89} and mainly the parallel momentum increases. Instead, for the majority of electrons the acceleration time is considerably longer than a few $\Oi^{-1}$, for some particles by a factor ten (e.g. \textit{particle 3}, Fig.~\ref{fig:accel_3}). The trajectories in $p_{\parallel} - p_{\perp}$ space show that most of the energization is associated with an increase of the \textit{perpendicular} momentum. It is interspersed with rapid pitch-angle scattering that is visible in $p_{\parallel} - p_{\perp}$ plots as arcs at constant total momentum, indicating elastic scattering that does not by itself provide energy gain.

The observed behavior has the characteristics of stochastic shock drift acceleration (SSDA), recently described in \citet{Katou-19}. In this mechanism particles are confined at the shock by stochastic pitch-angle scattering off magnetic turbulence, while gaining energy through SDA. The extended interaction time with the shock increases the total energy gain and hence provides more efficient acceleration than standard SDA or DSA. The presence of the \textit{multi-scale} turbulence in the shock is essential for electron acceleration to high energies in the SSDA process, which we demonstrate below.

As described in Section~\ref{sec:evolution}, turbulence on a variety of scales appears at the overshoots after the emergence of the rippling modes at $t\gtrsim 25\Oi^{-1}$. At earlier times single-cycle SDA operate, an example of which is the first shock encounter by \textit{particle~2} at $t\Oi\approx 10-12$. However, already in the laminar shock phase SDA is modified by particle scattering. This is well illustrated for \textit{particle~1}, whose acceleration at $t\Oi\approx 12-18$ involves intervals of either parallel or perpendicular momentum gains, and only in the final phase, $t\Oi\approx 18.5-20.5$, the energy accrual is mainly seen in $p_{\parallel}$, indicating pure SDA. In effect,
the interaction time of \textit{particle~1}, $t_{\mathrm{acc}} \approx 9.5\Oi^{-1}$, is much longer, and
its final energy,  $(\gamma - 1) = 5$, is considerably higher than theoretically expected for this particle $(\gamma - 1)_{\mathrm{SDA}} \approx 2.5$. It is even somewhat higher than expected maximum for standard SDA, $(\gamma - 1)_{\mathrm{SDA}}^{\mathrm{max}} \approx 4.5$, at the simulated conditions.
Local wave-particle interactions modify the upstream electron spectra that in the earlier, laminar phase were consistent with energization by single-cycle SDA.
We suppose that electron scattering in this phase is provided by whistlers at the overshoot.

Particle scattering during the laminar shock phase provides only a minor enhancement in the acceleration efficiency compared to pure SDA. Electrons that arrive at the rippled shock interact with wide-band turbulence, including the long-wave ripple modes at the overshoot, which allows the electrons to gain higher energies. \textit{Particle~2} achieves $\gamma \approx 9$ at its second shock encounter lasting $6\,\Oi^{-1}$ (Fig.~\ref{fig:accel}), and \textit{particle~3} finds even better scattering conditions and reaches $\gamma\approx 40$. The rate of the energy gain is not constant but grows as long as a particle resides at the shock and its energy increases (see \textit{top} panels of Figs.~\ref{fig:accel} and~\ref{fig:accel_3}).
At intervals of the fastest energy gain (at $t\Oi\approx 33-37$ for \textit{particle~2} and $t\Oi\approx 42-49$ for \textit{particle~3}), the momentum diagrams show a mixture of long arcs (pitch-angle scattering) and sequences of rapid small-angle scattering, during which $p_{\perp}$ grows. Particle-wave interactions with broad-band turbulence, including the largest-scale waves present in the shock, are thus vital for electron energization at this stage.

We demonstrated in Section~\ref{sec:evolution} that multi-scale turbulence is present in a wide region harboring overshoots and undershoots, suggesting that electrons may be accelerated behind the shock front. \textit{Particles 4} and \textit{5} shown in Figure~\ref{fig:accel-dwn} are examples of that. The majority of the particles that populate the high-energy tail in both the upstream and downstream spectra in fact gained their energy behind the shock.

The majority of such particles is energized around the second overshoot. Pitch-angle scattering confines the electrons between the undershoot and overshoot, where they tap energy in the weak motional electric field that persists there. Particles are either picked-up from the downstream population (\textit{particle 4}) or transmitted from the upstream (\textit{particle 5}). They first interact with short-wave whistlers and/or medium-scale ripples present at the second overshoot. The latter cascade towards longer wavelengths, providing conditions for continuous resonant electron scattering. The $p_{\parallel} - p_{\perp}$ phase-space plots in Figure~\ref{fig:accel-dwn} demonstrate strong scattering. It can at times involve multiple reflections between the first and the second overshoot (e.g., at $t\Oi\approx 19.5-25$  and $t\Oi\approx 30-35$ for \textit{particle~4}), that are possible if the phases of an electron orbit and the scattering centers match in both overshoots. The energy gains mainly arise in $p_{\perp}$.


The acceleration can continue for some electrons that return back to the shock surface and further resonantly interact with long-wave ripples there (e.g., \textit{particle 4}). This results in very high electron energies. Energetic electrons are observed to escape upstream at a higher rate than in the laminar shock stage. EFI-induced waves are then further amplified in the upstream region, providing additional confinement for particles undergoing SSDA at the shock front.

The features of electron acceleration described above account for the temporal evolution in the upstream electron spectra. Wide-range non-thermal tails first comprise but a few particles that find favorable conditions for acceleration in emerging multi-scale turbulence. They form a low-density high-energy population in the spectrum (see Section~\ref{subsec:distr-upstr}). The number of energetic particles quickly increases once the rippling modes are established.
The maximum electron energy is set by the condition, that the particle gyro-radius may not exceed the wavelength of the scattering turbulence.
For the ripples, $\lambda_{\mathrm{rippl}}\approx 16\lsi$, this condition limits electrons to energies $\gamma\approx 40$, which is roughly consistent with the observed cut-off in the upstream spectra.
\rev{Some particles are still accelerated to even higher energies, finding favorable scattering conditions. However, they escape upstream due to the absence of long-wave turbulence that can confine them at the shock. This explains a disappearance at $t\approx t_{\mathrm{max}}$ of low-density highest-energy particle population (compare a  
drop in the maximum Lorentz factor from $\gamma_{\mathrm{max}}\approx 60$ to $\gamma_{\mathrm{max}}\approx 30$ at cutoff level of $10^{-6}$ in Fig.~\ref{fig:spectra}b).}

Electron acceleration behind the shock front also explains the form of the downstream spectra. As we stated above, non-thermal components in downstream spectra are composed of particles undergoing SSDA behind the shock front and subsequently escaping downstream, for example \textit{particle 5}. In fact, this spectral component is formed long before any substantial number of upstream-accelerated electrons could be advected through the shock. Power-law tails also develop much earlier than in the upstream region, on account of efficient electron scattering off whistlers and small-scale ripples at the second overshoot, that appear already in the laminar shock phase (see Section~\ref{subsec:structure}).

\begin{figure}
\centering
\includegraphics[width=0.99 \linewidth, clip]{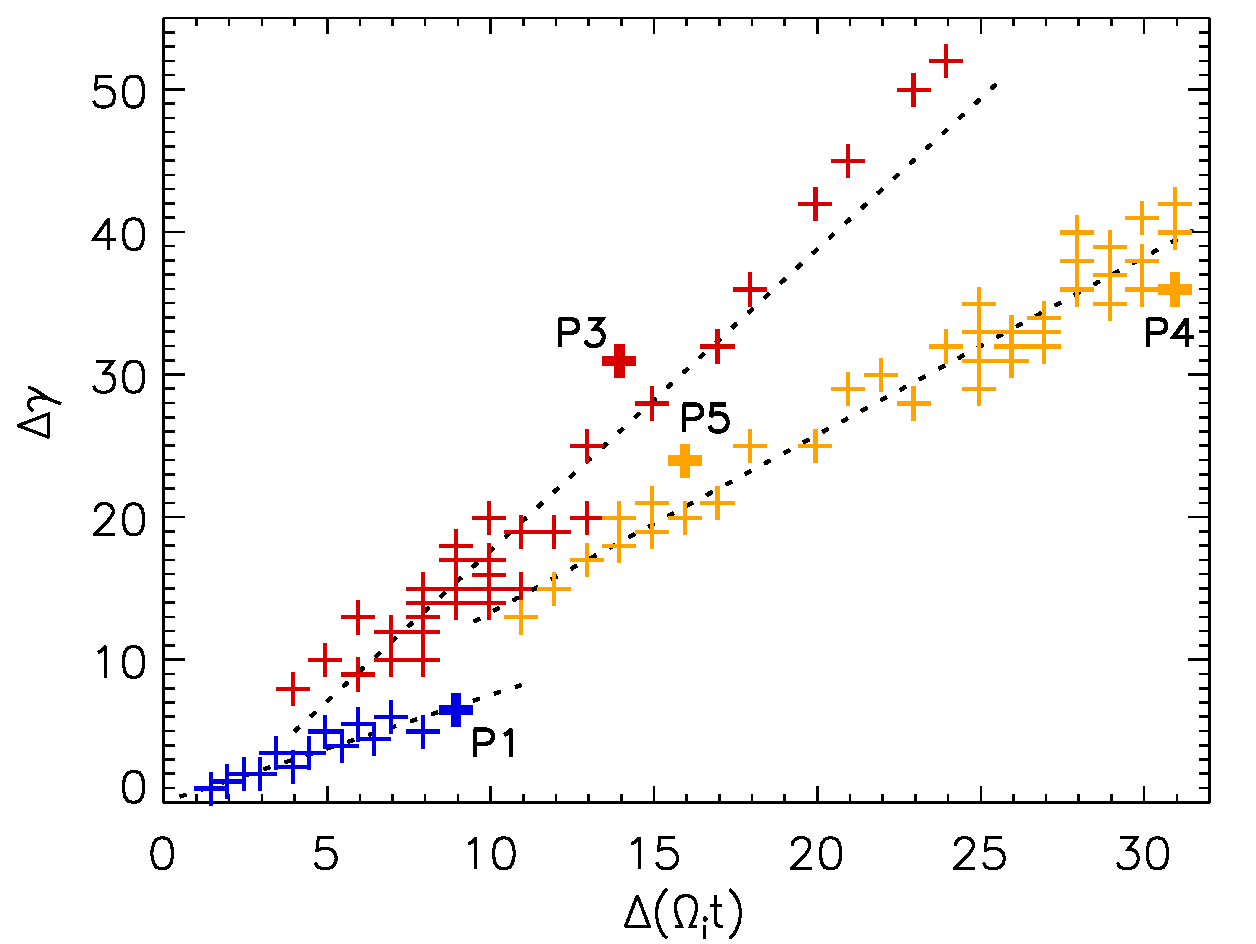}
\caption{Electron energy gain versus the confinement time for three different conditions: at the laminar shock (\textit{blue}), at the rippled shock (\textit{red}), and at the second overshoot (\textit{yellow}). The plus signs refer to individual particles, and those shown in Figs.~\ref{fig:accel-multi} and  \ref{fig:accel}~-~\ref{fig:accel-dwn} are marked as P1, P3, P4, and P5. Linear regression yields the energy-gain rates, $d \gamma / d(\Oi t)$, that are indicated by dotted lines. The values are around $ 0.75$, $ 2.1$, and $1.25$, respectively.}
\label{fig:accel-rate}
\end{figure}

Figure ~\ref{fig:accel-rate} demonstrates the efficiency of electron acceleration for the different conditions. Each plus sign stands for an individual traced particle, and the color distinguishes particles accelerated at the laminar shock (\textit{blue}), at the rippled shock front (\textit{red}), and behind the shock in the region around the second overshoot (\textit{yellow}). Each group roughly follows its own linear trend, implying a constant energy-gain rate. For the laminar shock we find $d \gamma / d(\Oi t) \approx 0.75$, whereas at the rippled shock the rate is three times as high. For the particles accelerated at the second overshoot, which is also rippled, the interaction time is longer, $\Delta (\Oi t) \gtrsim 30$, and the energy-gain rate is intermediate, $d \gamma / d(\Oi t) \approx 1.25$. Thus, electron energization at a rippled shock is more efficient than that at a laminar shock by the factor of a two or three.

From Eq.~\ref{eq:SDA} the acceleration rate can be derived as 
\begin{equation}
\left( \frac{d \gamma}{d \Oi t} \right)_{\mathrm{drift}} = \frac{m_\mathrm{i}}{m_\mathrm{e}} \frac{v_\mathrm{sh}}{c}\frac{ v_{z}}{c},
\label{eq:accel}
\end{equation}
where we write the upstream motional electric field in units of the ion gyro frequency, $\Oi$.
Linear acceleration in z-direction would suggest $v_z\approx c$ and hence an energy-gain rate $d \gamma / d(\Oi t)\gtrsim 10$. In fact, our tracing data indicate $v_z\lesssim 0.1 c$ and hence slow energization, implying that gyration prevents linear acceleration. A similar restriction was identified for shock-surfing acceleration at perpendicular shocks \citep[see the appendix]{Bohdan-19}. The observed rates of the energy gain also reflect variations in the amplitude of the large-scale electric field at the shock. Particle energization at the laminar shock involves drift along the shock ramp, in which the motional electric field drops from its upstream value, $E_{0z}$, to zero at the first overshoot. Our simulations show that the average amplitude of $E_{z}$ in this region is $E_{z}\approx E_{0z}/2$. Combined with the low out-of-plane speed, $v_z/c\lesssim 0.1$, the estimated energy-gain rate, $d\gamma / d(\Oi t) \approx 0.5$, is close to the observed one in the laminar shock.
Above a certain energy the Larmor radius is large enough to extend the gyration into the upstream region, where $E_{z}$ is stronger. This can enhance the energy-gain rate, visible, e.g., at $\Oi t \approx 18$ in Figure~\ref{fig:accel}a.
Electrons are accelerated at a rippled shock to much higher energies and at certain stage can probe the entire $E_{0z}$ during their drift. An example is \textit{particle 3} around $\Oi t \approx 42$ in Figure~\ref{fig:accel_3}. The $v_z$ speed then oscillates with amplitude that is a substantial fraction of $c$, but the average drift is much slower, $v_z/c\approx 0.2$. The energy-gain rate increases with time, so that the resulting rate, $d\gamma / d(\Oi t) \approx 2.1$, is an average over the entire acceleration time, during which particles probe different values of the motional electric field and their mean $v_z$ steadily grows. 
In contrast, particles accelerated behind the shock (Fig.~\ref{fig:accel-dwn}) for the most time probe the motional electric field in the undershoot, whose average amplitude is $\sim E_{0z}/2$. This can explain the  twice smaller energy gain rate compared to electrons accelerated at the rippled shock.

\section{Summary and discussion}
\label{sec:sum}

We investigate the conditions necessary for electron injection into DSA at merger shocks in a hot intracluster medium.
For that purpose, we performed a large-scale and long-duration 2D3V PIC simulation of a quasi-perpendicular shock of low Mach number, $M_s=3$, that propagates in plasma with a high plasma beta, $\beta=5$.  Our simulation resolves, and evolves to their nonlinear development, both the electron-scale and the ion-scale structures, the latter including corrugations of the shock front. 

Earlier studies of essentially laminar shocks indicated multi-cycle SDA providing electron pre-acceleration. This process relies on the presence of EFI waves in the upstream region that scatter SDA-reflected electrons back to the shock for repeated interactions. For our setup of a subluminal shock with $\theta_{\mathrm{Bn}} = 75^{\circ}$, EFI waves should be weakly driven, and so multi-cycle SDA should be inefficient. 
On the contrary, we observe numerous efficiently energized electrons whose spectra feature very extended non-thermal tails both upstream and downstream of the shock. The accelerated particles are produced through stochastic SDA, a process in which electrons are confined at the shock by pitch-angle scattering off turbulence and gain energy from the motional electric field. SSDA already operates during the laminar shock phase, enhancing the energy gain of SDA, but when the shock ripples appear, the energization rate considerably increases further. Rippling and other ion-scale waves that are driven by effective ion temperature anisotropy, together with electron-scale waves, that are correspondingly excited through electron temperature anisotropy, provide multi-scale magnetic turbulence that is essential for SSDA as it ensures efficient pitch-angle scattering at all times.  

Electrons gain energy both at the shock front and in the near-downstream region extending to the second overshoot. The upstream spectra are built from particles that experienced SSDA at the shock front and from those that interacted with turbulence immediately downstream and were subsequently scattered back upstream. The maximum energy of upstream electrons is sufficient for their injection into DSA, on account of their large Larmor radii. However, DSA cannot be observed in our simulation because the computational box is too small and the simulation time too short to capture the driving of long-wave turbulence by upstream-streaming particles.

Spectral tails in the downstream region are primarily composed of electrons that were accelerated around the second overshoot. For these particles we demonstrate for the first time a power-law tail with index 
\rev{$p\approx 2.5$}, in agreement with observations. 
We show that the rate of the energy gain for particles accelerated at the rippled shock front is twice larger than for energization behind the shock and three times larger than for interaction with the laminar shock.

Observational evidence for electron injection via SSDA at the Earth's bow shock has been recently provided by the Magnetospheric Multiscale mission \citep{Amano-20}. Waves that diffusively confine electrons within the acceleration region were identified as high-frequency coherent whistlers with right-hand polarization \citep[see also][]{Oka-17}. In numerical experiments, SSDA was observed in fully-kinetic 3D PIC simulations of quasi-perpendicular high-Mach-number shocks of young supernova remnants \citep[$M_{\mathrm{s}} \gtrsim 20, \beta = 1$,][]{Matsumoto-17} and in hybrid PIC and test-particle studies of solar-wind shocks
\citep[$M_{\mathrm{s}} =6.6, \beta = 1$,][]{2019MNRAS.482.1154T}.
In supernova remnant shocks the stochasticity is provided by Weibel modes at the shock foot, whereas shock-surface fluctuations may be more relevant under solar-wind conditions. Here we show that at shocks driven by galaxy mergers, electron scattering is due to multi-scale turbulence in the entire shock transition.

The generation of multi-scale turbulence, including AIC-driven waves, has been recently confirmed for supercritical shocks and high plasma beta, $\beta=20-100$ \citep{Ha-21}. They showed that SSDA is responsible for the majority of the most energetic electrons in the upstream region and that it can significantly contribute to electron pre-acceleration. 
The pre-acceleration efficiency depends only weakly on the plasma beta for their fiducial magnetic-field obliquity angle, $\theta_{\mathrm{Bn}} = 63^{\circ}$, and increases with $\theta_{\mathrm{Bn}}$, provided that is below ${\theta_{\rm limit}}$ to ensure efficient EFI wave generation. Our result obtained for $\theta_{\mathrm{Bn}}>\theta_{\rm limit}$ suggests that SSDA alone can provide electron injection into DSA. Verification of this supposition with simulations in a range of magnetic-field obliquities will be a subject of future studies.  


\section*{Acknowledgements}

This work has been supported by Narodowe Centrum Nauki through research projects DEC-2013/10/E/ST9/00662 (O.K.,J.N.), UMO-2016/22/E/ST9/00061 (O.K.) and 2019/33/B/ST9/02569 (J.N.). This research was supported by PLGrid Infrastructure. Numerical experiments were conducted on the Prometheus system at ACC Cyfronet AGH and also on resources provided by The North German Supercomputing Alliance (HLRN) under projects bbp00003, bbp00014, and bbp00033. This work was also supported by JSPS-PAN Bilateral Joint Research Project Grant Number 180500000671.

\end{document}